\documentclass[%
 reprint,
superscriptaddress,
 amsmath,amssymb,
 aps,
pra,
twocolumn,
floatfix,
]{revtex4-1}

\usepackage[english]{babel}
\usepackage{graphicx}% Include figure files
\usepackage{dcolumn}% Align table columns on decimal point
\usepackage{bm}% bold math
\usepackage[caption=false]{subfig}
\usepackage{subdepth}
\usepackage{hyperref}% add hypertext capabilities
\usepackage{mathtools}
\usepackage[dvipsnames]{xcolor}

\begin{document}

\preprint{APS/123-QED}

\title{Two-dimensional phase-space picture of the photonic crystal Fano laser}% Force line breaks with \\

\author{Piotr M. Kami\'nski}
  \email{pmarka@elektro.dtu.dk}
  \affiliation{
Department of Electrical Engineering, Technical University of Denmark, {\O}rsteds Plads, 2800 Kongens Lyngby, Denmark\\
}
\author{Samel Arslanagi\'c}
\email{sar@elektro.dtu.dk}
  \affiliation{
Department of Electrical Engineering, Technical University of Denmark, {\O}rsteds Plads, 2800 Kongens Lyngby, Denmark\\
}
\author{Jesper M{\o}rk}
\affiliation{
Department of Photonics Engineering, Technical University of Denmark, {\O}rsteds Plads, 2800 Kongens Lyngby, Denmark\\
}
\author{Jensen Li}
\email{jensenli@ust.hk}
\affiliation{
Department of Physics, Hong Kong University of Science and Technology, Clear Water Bay, Hong Kong, China\\
}

\date{\today}% It is always \today, today,

\begin{abstract}
The recently realized photonic crystal Fano laser constitutes the first demonstration of passive pulse generation in nanolasers [Nat. Photonics $\boldsymbol{11}$, 81-84 (2017)]. We show that the laser operation is confined to only two degrees-of-freedom after the initial transition stage. We show that the original 5D dynamic model can be reduced to a 1D model in a narrow region of the parameter space and it evolves into a 2D model after the exceptional point, where the eigenvalues transition from being purely to a complex conjugate pair. The 2D reduced model allows us to establish an effective band structure for the eigenvalue problem of the stability matrix to explain the laser dynamics. The reduced model is used to associate a previously unknown origin of instability with a new unstable periodic orbit separating the stable steady-state from the stable periodic orbit.
\end{abstract}

\maketitle

%\tableofcontents

\section{\label{sec:intro}Introduction}

Integrated photonic circuits require energy efficient, fast and compact light sources \cite{miller_device_2009}. Particularly promising candidates to realize them are photonic crystal (PhC) lasers due to their flexibility in design and precise control of the cavity properties \cite{akahane_high-q_2003-1, tran_directive_2009}. PhC lasers can be electrically driven and allow for modulation in the GHz range \cite{matsuo_ultralow_2013,jang_sub-microwatt_2015}. Moreover, they have been shown to exhibit very rich dynamics, e.g. spontaneous symmetry breaking \cite{hamel_spontaneous_2015}. Recently, a new type of PhC laser has been proposed \cite{mork_photonic_2014} where one of the mirrors arises due to a Fano resonance \cite{fano_effects_1961,limonov_fano_2017}. Furthermore, this laser has been demonstrated to be able to generate a self-sustained train of pulses at GHz frequencies, a property that has been observed only in macroscopic lasers thus far \cite{yu_demonstration_2017}. Generation of pulses by an ultracompact laser is of interest for applications in future on-chip optical signal processing.

\begin{figure}
\includegraphics[width=0.99\linewidth]{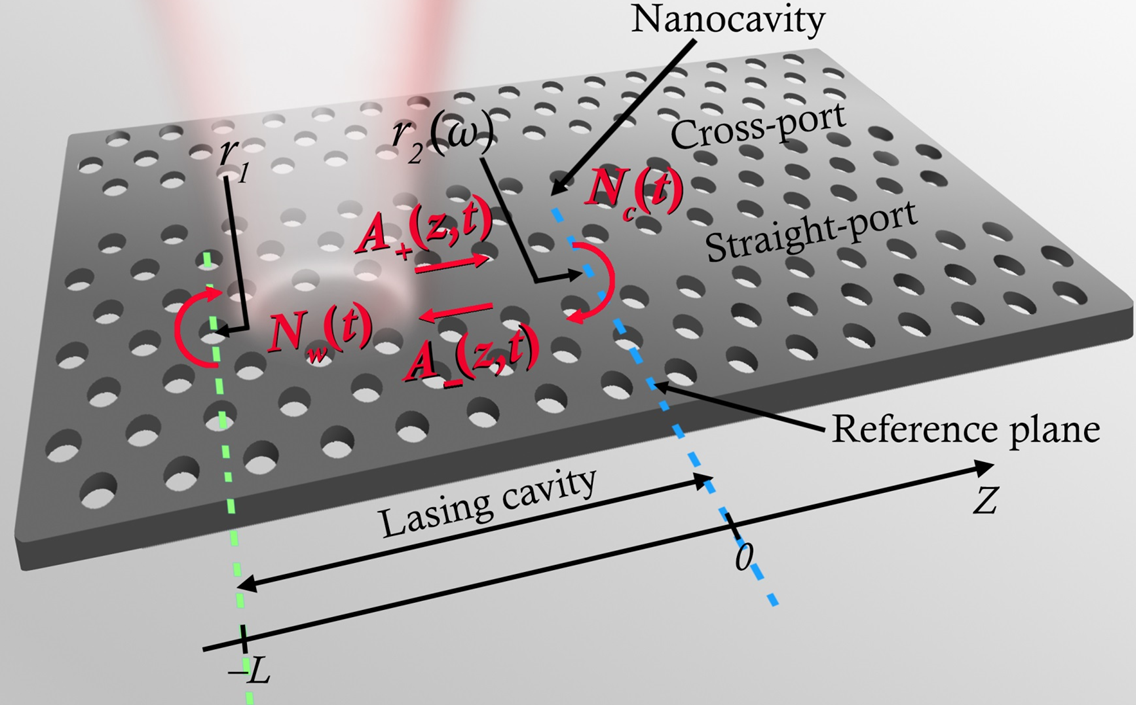}
\caption{Schematic of a PhC Fano laser. The active material is uniformly incorporated in the PhC slab. The lasing cavity is composed of a PhC line-defect waveguide terminated with a broad-band mirror (green) and a narrow-band mirror due to the Fano interference between the nanocavity and the waveguide. The length of the lasing cavity is defined as the distance between the broad-band mirror and a reference plane (blue). The dynamic variables are marked in red and are: the carrier densities in the waveguide and the nanocavity, $N_w(t)$ and $N_c(t)$, respectively, and the right and left propagating complex field envelopes $A_+(0, t)$ and $A_-(0, t)$, respectively, evaluated at the reference plane.}
\label{fig:Fano_conf}
\end{figure}

The configuration of the Fano laser is shown in Fig. \ref{fig:Fano_conf}. The active material may be composed of several layers of InAs quantum dots or quantum wells and is incorporated inside the InP PhC membrane. The laser cavity is composed of a PhC line-defect waveguide blocked with a PhC mirror on the left side forming a broad-band mirror, whereas the right mirror is due to the Fano interference between the nanocavity and the waveguide. The Fano resonance arises due to the interference of a discrete mode of the nanocavity with the continuum of PhC waveguide modes. The spectral width of the resonance is determined by the quality factor of the nanocavity enabling the realization of a narrow-band mirror. The dynamic operation of the laser is modeled using a combination of coupled-mode theory and conventional laser rate equations \cite{rasmussen_theory_2017}. The model has been used to demonstrate that there are two regimes of operation, the continuous-wave regime and a self-pulsing regime \cite{rasmussen_theory_2017}. Particularly, it has been shown that as the real part of any of the eigenvalues of the underlying stability matrix, evaluated at the steady-state, becomes positive, the relaxation oscillation becomes undamped resulting in the laser becoming unstable and self-pulsing behaviour setting in \cite{rasmussen_theory_2017,rasmussen_modes_2018}. However, it does not fully explain the origin of instability in the whole parameter space of the laser as there exists a region in which the laser can become unstable even when all the steady-state eigenvalues are negative \cite{rasmussen_theory_2017}.

The purpose of this work is to analyze not only the steady-state eigenvalues of the stability matrix of the dynamic model, but also the instantaneous eigenvalues during the laser operation. Moreover, we determine the 'minimal' model for the laser that is required to explain the dynamics in different regimes, thereby obtaining an alternative perspective on the dynamics of the Fano laser. We thus demonstrate that the laser operation can be effectively modeled by a one-dimensional (1D) system of differential equations in a limited region of the parameter space when the steady-state eigenvalues are purely real and that it evolves into an effective two-dimensional (2D) system beyond the steady-state exceptional point, when the eigenvalues form complex conjugate pairs. These findings are used to determine the origin of the instability that is observed when the steady-state eigenvalues are negative. We notice that the analysis of instabilities and chaos in injection-locked lasers, e.g., using bifurcation analysis has been very successful \cite{mork_chaos_1992,krauskopf_bifurcation_2000,wieczorek_dynamical_2005,erzgraber_bifurcation_2007}. Here, we use it to analyse origin of laser instability when the steady-state is stable and set an important goal to identify reduced systems for getting further physical insight.

The manuscript is organized as follows. In Sec. \ref{sec:dyn_mod} we introduce the model used to describe the laser dynamics. In Sec. \ref{sec:2d_pict} we show that the laser operation can be understood by means of a 2D phase-space picture and we analyze the steady-state and instantaneous eigenvalues of the stability matrix. In Sec. \ref{sec:Hopf} we exploit the simplified 2D model to associate a self-pulsing operation, when a steady-state is stable to a generalized Hopf (Bautin) bifurcation which is characterized by a bifurcation of two periodic orbits and an equilibrium point (steady-state) \cite{govaerts_numerical_2000,kuznetsov_elements_2004}.

\section{\label{sec:dyn_mod}Dynamic model of the Fano laser}

We next briefly describe the procedure required to establish the dynamic model of the Fano laser; for more details refer to \cite{rasmussen_theory_2017}. The complex field is decomposed into the fields propagating to the right and left from the reference plane, see Fig. \ref{fig:Fano_conf}. By combining the boundary conditions for both fields, we can arrive at the oscillation condition \cite{mork_photonic_2014}:

\begin{equation}
r_1(\omega_s)r_R(\omega_s,\omega_c)e^{i2k(\omega_s,N_s)L}=1,
\label{eq:osc_cond}
\end{equation}

\noindent where $r_1$ and $r_R$ are the broad-band (left) and the narrow-band (right) reflection coefficients, respectively. $r_R$ is determined using the coupled-mode theory \cite{fan_temporal_2003,wonjoo_suh_temporal_2004,kristensen_theory_2017}, while $r_1$ is the reflection coefficient due to the PhC band gap and has to be transformed towards the common reference plane using standard transmission line theory \cite{tromborg_transmission_1987}. $k$ is the complex wavenumber of the waveguide, $L$ is the length of the lasing cavity, and $\omega_c$ is the resonance frequency of the nanocavity. The condition in Eq. (\ref{eq:osc_cond}), is solved for $(\omega_s,N_s)$, which are the steady-state lasing frequency and carrier density, respectively. They serve as expansion points of the dynamic model. There are multiple solutions of Eq. (\ref{eq:osc_cond}) \cite{mork_photonic_2014, rasmussen_modes_2018} among which the one with the lowest modal threshold gain is chosen. The wavenumber, $k$, accounts for dispersion of the refractive index of the PhC membrane and the gain of the active material.

Subsequently, the boundary condition is solved for the left propagating field and then the term $1/r_1(\omega)e^{i2k(\omega_s,N_s)L}$ is Taylor expanded around the steady-state operation point $(\omega_s,N_s)$ and a first-order differential equation for the right-propagating complex field envelope evaluated at the reference plane $A_+(t)$ is derived using the Fourier transform. In the special case of an open waveguide considered here, the coupled-mode equation for the field in the nanocavity can be directly reformulated as an equation for the left-propagating complex field envelope evaluated at the reference plane $A_-(t)$. The equations for $A_+(t)$ and $A_-(t)$ are complemented with the traditional rate equations for carrier densities in the waveguide and the nanocavity.

Since the variables introduced above differ by orders of magnitude, we introduce dimensionless near-unity variables in order to improve numerical stability. Moreover, detuning from the expansion point frequency $\omega_s$ results in time-varying real and imaginary parts of $A_+(t)$ and $A_-(t)$ at the steady-state. Because of that, the differential equations for $A_+(t)$ and $A_-(t)$ are separated into equations for amplitudes and phase evolutions by the following substitution: $A_+(t)=A_{0+}|a_+|(\tau/(G_{NC}N_0))e^{i\phi_+(\tau/(G_{NC}N_0)}$, $A_-(t)=A_{0-}|a_-|(\tau/(G_{NC}N_0))e^{i\phi_-(\tau/(G_{NC}N_0)}$, where $A_{0+}$ and $A_{0-}$ are the normalization constants, $a_-(\tau)$ and $a_+(\tau)$ are the normalized complex field envelopes, $\tau=t G_{NC}N_0$ is the normalized time, $G_{NC}=\Gamma_Cv_gg_N$, and $v_g$ is the group velocity. The system depends solely on the phase difference $\Delta\phi(\tau)=\phi_-(\tau)-\phi_+(\tau)$; thus by subtracting the equations for phase evolutions $\phi_+(\tau)$, $\phi_-(\tau)$ and exploiting linearity of differentiation, these equations can be combined into one. This leads us to the following system of five differential equations describing the dynamics of the laser:

\begin{widetext}
\begin{subequations}\label{eq:set_diff_eqs}
\begin{equation}
\frac{d|a_+|(\tau)}{d\tau}=-\frac{\gamma_L|a_+|(\tau)}{G_{NC}N_0}+\frac{\Gamma|a_+|(\tau)(n_w(\tau)-n_s)}{2\Gamma_C}+\frac{\gamma_L}{G_{NC}N_0}|a_+|(\tau)\mathrm{Re}\bigg(\frac{A_-(\tau)}{r_RA_+(\tau)}\bigg)
\end{equation}
\begin{equation}
\frac{d|a_-|(\tau)}{d\tau}=-\frac{P\gamma_C}{G_{NC}N_0}|a_-|(\tau)\mathrm{Re}\bigg(\frac{A_+(\tau)}{A_-(\tau)}\bigg)-\frac{\gamma_T|a_-|(\tau)}{G_{NC}N_0}+\frac{|a_-|(\tau)(n_c(\tau)-1)}{2}
\end{equation}
\begin{equation}
\frac{d\Delta\phi(\tau)}{d\tau}=-\frac{\alpha}{2}(n_c(\tau)-1)+\frac{\Gamma \alpha(n_w(\tau)-n_s)}{2\Gamma_C}-\frac{\Delta\omega}{G_{NC}N_0}-\mathrm{Im}\bigg(\frac{r_RP\gamma_C{A}^2_+(\tau)+\gamma_L{A}^2_-(\tau)}{r_RG_{NC}N_0A_-(\tau)A_+(\tau)}\bigg)
\end{equation}
\begin{equation}
\frac{dn_w(\tau)}{d\tau}=\frac{-|a_+(\tau)|^2(n_w(\tau)-1)-n_w(\tau)+j_c}{G_{NC}N_0\tau_s}
\end{equation}
\begin{equation}
\frac{dn_c(\tau)}{d\tau}=\frac{-|a_-(\tau)|^2(n_c(\tau)-1)-n_c(\tau)}{G_{NC}N_0\tau_c}
\end{equation}
\end{subequations}
\end{widetext}

Here, $n_w(\tau)$ and $n_c(\tau)$ are the carrier densities in the waveguide and the nanocavity, respectively, normalized with respect to the transparency carrier density, $N_0$. $\gamma_L=v_g/(2L)$ is the inverse of the cavity roundtrip time, $n_s$ is the steady-state carrier density obtained from the oscillation condition normalized with respect to $N_0$, $\Delta\omega=\omega_c-\omega_s$ is the detuning of the steady-state lasing frequency, $\omega_s$, from the cavity resonance frequency, $\omega_c$, and $j_c$ is the normalized effective pumping current, which includes the injection efficiency.

Subsequently, we linearize the problem by calculating the total derivative of Eqs. (\ref{eq:set_diff_eqs}) with respect to $\tau$. The system of equations describing the laser dynamics in Eqs. (\ref{eq:set_diff_eqs}) can be expressed in the short form as a function $V(\cdot)$ of the state vector $\vec{\psi}(\tau)$:

\begin{subequations}\label{eq:interpretation}
\begin{equation}\label{eq:position}
\vec{\psi}(\tau)=\{|a_+(\tau)|, |a_-(\tau)|, \Delta\phi(\tau), n_w(\tau), n_c(\tau)\}
\end{equation}
\begin{equation}\label{eq:velocity}
  \frac{d\vec{\psi}(\tau)}{d\tau}=V(\vec{\psi}(\tau))
\end{equation}
\end{subequations}

By taking the total derivative of $V(\vec{\psi}(\tau))$, we obtain a directional derivative along the curve parameterized by $\tau$:

\begin{equation}\label{eq:acceleration}
\frac{d^2\vec{\psi}(\tau)}{d\tau^2}=\nabla_{\vec{\psi}} V(\vec{\psi}(\tau))\frac{d\vec{\psi}(\tau)}{d\tau}=\boldsymbol{A}(\vec{\psi}(\tau))\frac{d\vec{\psi}(\tau)}{d\tau}
\end{equation}

Consequently, $d\vec{\psi}(\tau)/d\tau$ in Eq. (\ref{eq:velocity}) is interpreted as the velocity of the state vector and is expressed as a function of the current position of the state vector in Eq. (\ref{eq:set_diff_eqs}). $d^2\vec{\psi}(\tau)/d\tau^2$ is interpreted as the acceleration of the state vector, see Eq. (\ref{eq:acceleration}). Matrix $\boldsymbol{A}$ is the so-called Jacobian matrix; its eigenvalues $\lambda$ are used to determine the stability of the laser when evaluated at the steady-state. The system is stable if all eigenvalues have negative real parts. On the other hand, if any eigenvalue has a positive real part, the system is unstable. The matrix $\boldsymbol{A}$ is purely real, but not symmetric as we separated the complex field envelopes into the magnitudes $|a_+(\tau)|$, $|a_-(\tau)|$ and the phase difference $\Delta\phi$. Therefore, the matrix is non-Hermitian and we have to distinguish between right $\vec{v}$ and left $\vec{w}$ eigenvectors which are normalized so that $\boldsymbol{W}^T\boldsymbol{V}=\boldsymbol{I}$ is satisfied \cite{morse_methods_1953, ibanez_adiabaticity_2014}. The columns of $\boldsymbol{W}$ and $\boldsymbol{V}$ are the left and the right eigenvectors, and $\boldsymbol{I}$ is the identity matrix. Furthermore, eigenvalues of the matrix $\boldsymbol{A}$ can be purely real or form complex conjugate pairs \cite{arfken_mathematical_2005}. In the following sections, we use Eqs. (\ref{eq:set_diff_eqs}) and (\ref{eq:interpretation}) to investigate the origin of instability in case of a stable steady-state and to show that the original laser model can be simplified to a system of two differential equations.

\section{\label{sec:2d_pict}Two dimensional phase-space picture}

\subsection{Steady-state eigenvalues}

\begin{figure}
 \centering
  \subfloat[]{     \includegraphics[width=0.925\linewidth]{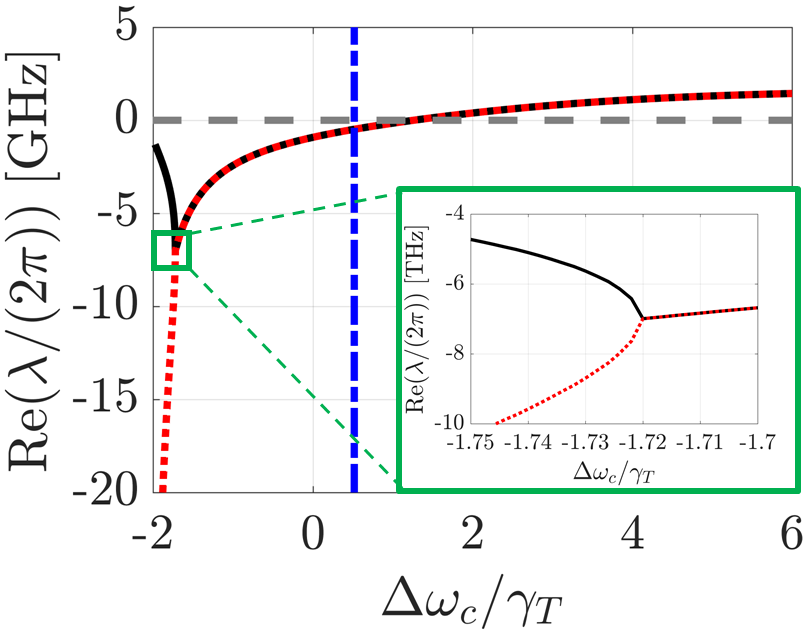}}\hfill
  \subfloat[]{     \includegraphics[width=0.925\linewidth]{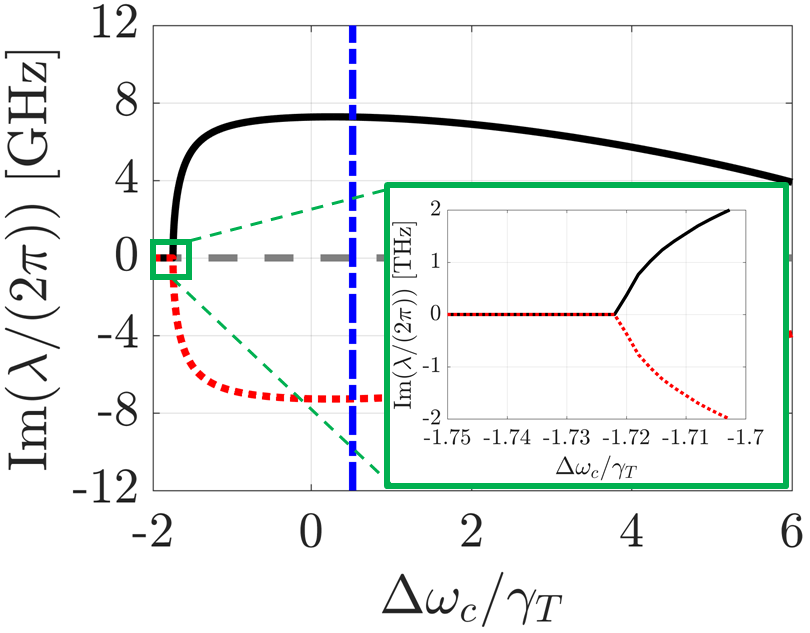}}
  \caption{Real a) and imaginary b) parts of the two steady-state eigenvalues of the matrix, $\boldsymbol{A}$ with the largest real parts. The horizontal dashed gray line indicates zero. The vertical dot-dashed blue line indicates $\Delta\omega_c$ for which Fig. \ref{fig:unit_cellphc} is obtained. The pumping current is set to $J=1.2 J_{thr}$. The green frame marks the position of the exceptional point, while the insets show the eigenvalues in its vicinity.}
  \label{fig:steady_state_eig}
\end{figure}

Above the threshold, the laser can exhibit two types of operation, the continuous wave and the self-pulsing operation \cite{rasmussen_theory_2017, yu_demonstration_2017}. Figure \ref{fig:steady_state_eig} shows the real and imaginary parts of the two steady-state eigenvalues of the Jacobian matrix, $\boldsymbol{A}$ with the largest real parts plotted versus $\Delta\omega_c$, which is the detuning of the cavity resonance frequency, $\omega_c$ from the resonance frequency of the isolated cavity, normalized with respect to $\gamma_T$. It is noted that $\Delta\omega_c$ defines $\Delta\omega$ in Eq. (\ref{eq:set_diff_eqs}) through the oscillation condition, Eq. (\ref{eq:osc_cond}), and is controlled externally. As our case study, we choose $\Delta\omega_c$ marked with the blue line in Fig. \ref{fig:steady_state_eig}.

Interestingly, it has been observed in \cite{rasmussen_theory_2017} that in the vicinity of the $\Delta\omega_c$ marked by the blue dashed line in Fig. \ref{fig:steady_state_eig} the laser can exhibit the continuous wave or the self-pulsing operation depending on the initial condition despite its steady-state eigenvalues having negative real parts and thus suggesting stable operation of the laser. However, origin of this instability has not been explained and is examined in Sec. \ref{sec:Hopf}. On the other hand, when $\Delta\omega_c$ is increased beyond $1.26\gamma_T$, the real parts become positive, the relaxation oscillation becomes undamped, the laser becomes unstable, and the state approaches a stable periodic orbit for any initial condition \cite{rasmussen_theory_2017}. All the following figures are obtained for the parameters listed in Table \ref{tab:table1}, while the pumping current is set to $J=1.2J_{thr}$, where $J_{thr}$ is the minimum threshold current.

\begin{table} %The best place to locate the table environment is directly after its first reference in text
\caption{\label{tab:table1}%
Laser parameters used in all numerical simulations.
}
\renewcommand{\arraystretch}{1.3}
\begin{ruledtabular}
\begin{tabular}{ccc}
\textrm{Parameter name} &
\textrm{Symbol} &
\textrm{Value}\\
\colrule
Transparency carrier density & $N_0$ & $1\cdot10^{24} \mathrm{m^{-3}}$ \\
Parity of the cavity mode & $P$ & $1$ \\
Linewidth enhancement factor & $\alpha$ & $1$ \\
Internal loss factor & $\alpha_i$ & $1000\mathrm{m^{-1}}$ \\
Lasing cavity length & $L$ & $5\mathrm{\mu m}$ \\
Carrier lifetimes & $\tau_s$, $\tau_c$ & $0.5$ns \\
Laser cavity volume & $V_{LC}$ & $1.05 \mathrm{\mu m^3}$ \\
Nanocavity volume & $V_{NC}$ & $0.243 \mathrm{\mu m^3}$ \\
Nanocavity resonance & $\lambda_r$ & $1554\mathrm{nm}$ \\
Reference refractive index & $n_{ref}$ & $3.5$ \\
Group refractive index & $n_{grp}$ & $3.5$ \\
Differential gain & $g_N$ & $5\cdot10^{-20}\mathrm{m^2}$ \\
Waveguide confinement factor & $\Gamma$ & $0.5$ \\
Nanocavity confinement factor & $\Gamma_C$ & $0.3$ \\
Left mirror reflectivity & $R_1$ & $1$ \\
Nanocavity-waveguide coupling & $\gamma_C$ & $1.14 \mathrm{ps^{-1}}$ \\
Nanocavity total passive decay rate & $\gamma_T$ & $1.21 \mathrm{ps^{-1}}$ \\
\end{tabular}
\end{ruledtabular}
\end{table}

\subsection{Exceptional points}

It is interesting to observe in Fig. \ref{fig:steady_state_eig} that for $\Delta\omega_c$ lower than $-1.72\gamma_T$, the real part of the two eigenvalues split and the eigenvalues become purely real, see Fig. \ref{fig:steady_state_eig}b). At $\Delta\omega_c=-1.72\gamma_T$, the two eigenvalues coalesce and not only the eigenvalues are identical at this point, but so are the eigenvectors as well \cite{dembowski_experimental_2001,heiss_exceptional_2004,berry_physics_2004,heiss_physics_2012,liertzer_pump-induced_2012}. This constitutes an exceptional point which is also known as a symmetry breaking point for a non-Hermitian system \cite{bender_generalized_2002,ruter_observation_2010,feng_non-hermitian_2017,el-ganainy_non-hermitian_2018}. However, exceptional points are a general phenomenon observed in optical waveguides \cite{klaiman_visualization_2008}, unstable laser resonators \cite{berry_mode_nodate}, coupled PhC nanolasers \cite{kim_direct_2016}, quantum systems \cite{lefebvre_resonance_2009}, electronic circuits \cite{stehmann_observation_2004} and mechanical resonators \cite{xu_topological_2016}. They only require non-Hermiticity of the system for their existence \cite{heiss_repulsion_2000,berry_physics_2004,heiss_exceptional_2004}. We emphasize that exceptional points may arise upon coalescence of eigenvectors/eigenvalues of any matrix, e.g. a Hamiltonian matrix \cite{rotter_non-hermitian_2009}, a S-parameter matrix \cite{chong_symmetry_2012} and an impedance/admittance matrices \cite{hanson_exceptional_2004}, to name a few. Exceptional points have also been linked to a self-pulsing mechanism in distributed feedback lasers \cite{bandelow_theory_1993, wenzel_mechanisms_1996} in which case the self-pulsing mechanism was attributed to dispersive quality factor self-switching similarly as in the case of the Fano laser \cite{yu_demonstration_2017}. In the present case, exceptional points arise due to dissimilar decay rates, $\gamma_C$, $\gamma_L$ and phenomenologically introduced gain terms $|a(\tau)|(n(\tau)-1)$, Eqs. (\ref{eq:set_diff_eqs}a) and (\ref{eq:set_diff_eqs}b). They play an analogous role to the loss and gain usually introduced as an imaginary part of the refractive index in parity-time symmetric systems \cite{feng_single-mode_2014,hodaei_parity-time-symmetric_2014}.

\subsection{Two-dimensional phase-space}

\begin{figure}
   \subfloat[]{\includegraphics[width=0.95\linewidth]{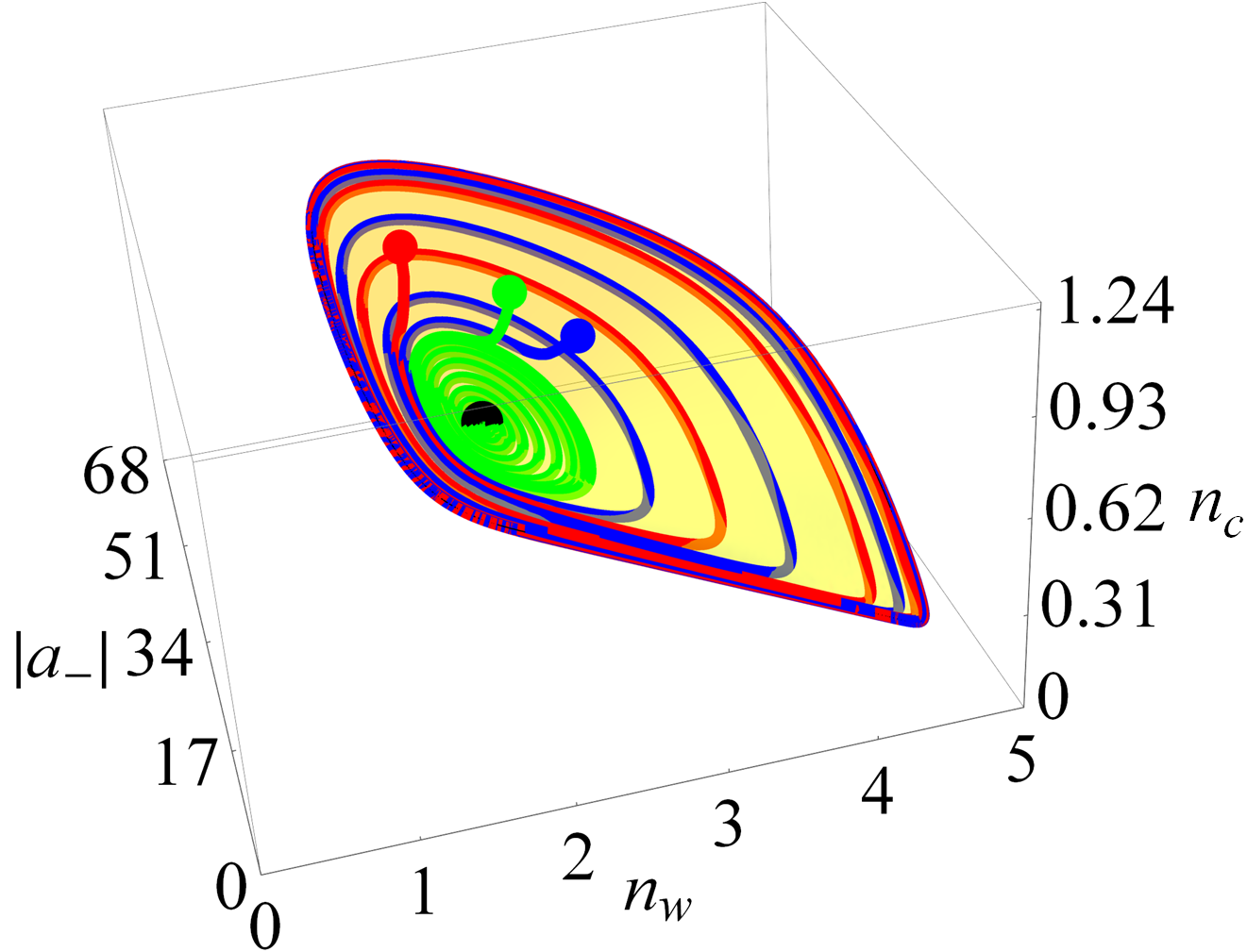}} \centering\\
   \subfloat[]{\includegraphics[width=0.925\linewidth]{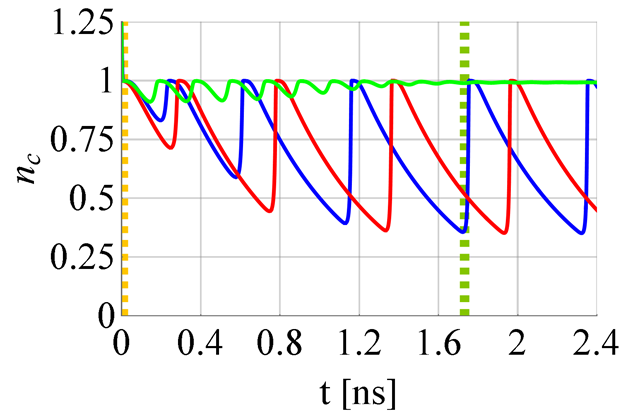}}\centering
  \caption{a) The trajectories of $n_c$ against $n_w$ and $|a_-|$. The black dot represents the steady state. The red, green and blue dots are the three different initial conditions lying above the yellow surface. During an initial transition stage, the state decays towards the surface. Then the state continues its evolution on the surface. b) The trajectories in time-domain show the initial transition stage lasting a few ps, the later transition stage lasting $\sim$$1.2$ns, and finally the self-pulsing stage at the edge of the surface. The vertical orange/green dashed lines mark the end of the initial/later transition stage.}
  \label{fig:unit_cellphc}
\end{figure}

In Fig. \ref{fig:unit_cellphc} we plot three trajectories of $n_c$, marked in red, green and blue versus $n_w$, $|a_-|$ and obtained for the three different initial conditions. The trajectories are parameterized by $\tau$. It is found that there are actually three different stages of the laser operation: the initial transition stage, the later transition stage and the self-pulsing stage. This is in contrast to the previously reported picture of two stages: the transition stage and the self-pulsing stage. The red, green and blue dots mark the initial conditions in Fig. \ref{fig:unit_cellphc}. It is found that at first they lie above a yellow surface, this is the initial transition stage which lasts only a few picoseconds. After a very short initial transition stage, the state reaches the surface at the time instant marked with the orange dashed line in Fig. \ref{fig:unit_cellphc}b). The state stays on the surface within the later transition stage and the self-pulsing/continuous wave stage. The later transition stage is when the state approaches the stable steady-state or the stable periodic orbit. Eventually, the state reaches the stable periodic orbit at the time instant marked with the green dashed line in Fig. \ref{fig:unit_cellphc}b). The state stays at the orbit unless perturbed; this stage is called the self-pulsing stage and takes place at the edge of the yellow surface.

Thus, it is found that once the state reaches the yellow surface, the state is confined to the surface. The phenomenon of data collapse to a surface happens also for the other two variables, $|a_+|$, $\Delta\phi$. Since the state always lies on the surface after a very short initial time, we conclude that two degrees of freedom are sufficient to specify the state after the initial transition stage and the propagation of the state is locally restricted to two directions. We note that this phenomenon is a general feature of a dynamical system close to a Hopf bifurcation and is called a reduction to the center manifold \cite{kuznetsov_elements_2004, seydel_practical_2010, guckenheimer_nonlinear_1983}. The dimension of the center manifold is strictly related to the number of steady-state eigenvalues, the real parts of which cross zero \cite{kuznetsov_elements_2004, seydel_practical_2010, guckenheimer_nonlinear_1983}. In Fig \ref{fig:steady_state_eig}a), we have seen that in the present case there are two eigenvalues with real parts crossing zero, while all the remaining eigenvalues have negative real parts giving rise to a stable manifold. Thus, the center manifold is two-dimensional as it is confirmed by the yellow curved surface in Fig. \ref{fig:unit_cellphc}a). The dynamics in the remaining three directions quickly approach the surface during the initial transition stage.

In Fig. \ref{fig:unit_cellphc}, $n_w$ and $|a_-|$ are the two degrees of freedom, while all the remaining degrees of freedom \{$|a_+|$, $\Delta\phi$, $n_c$\} of the state vector $\vec{\psi}$ are expressed as functions of the variables $n_w$ and $|a_-|$ after the initial transition stage:

\begin{subequations}\label{eq:interpretation}
\begin{align}\label{eq:position2}
  \vec{\psi}= &\{|a_+|(|a_-|, n_w), |a_-|, \Delta\phi(|a_-|, n_w), n_w, n_c(|a_-|, n_w)\},
\end{align}

Similarly, equations for each component of the velocity vector $d\vec{\psi}/d\tau$ in Eqs. (\ref{eq:set_diff_eqs}) and (\ref{eq:velocity}) can be expressed as functions of $n_w$ and $|a_-|$:

\begin{align}\label{eq:velocity2}
  \frac{d\vec{\psi}}{d\tau}=\{& V_1(|a_-|, n_w), V_2(|a_-|, n_w), V_3(|a_-|, n_w), \nonumber\\
  & V_4(|a_-|, n_w), V_5(|a_-|, n_w)\}.
\end{align}
\end{subequations}

By taking the total derivative of $d\vec{\psi}(\tau)/d\tau$, we obtain:

\begin{equation}\label{eq:acceleration2}
\bigg(\frac{d^2\vec{\psi}(\tau)}{d\tau^2}\bigg)_i=\bigg\{\frac{\partial V_i}{\partial |a_-|}\frac{d|a_-|}{d\tau}+\frac{\partial V_i}{\partial n_w}\frac{dn_w}{d\tau}\bigg\},
\end{equation}

\noindent which, when compared with Eq. (\ref{eq:acceleration}), indicates that the laser dynamics can be locally approximated by a $2\times2$ Jacobian matrix $\boldsymbol{A}$.

The functions of $n_w$ and $|a_-|$ in Eq. (\ref{eq:position2}) and the yellow surface in Fig. \ref{fig:unit_cellphc} are approximated by polynomials. In order to do that, we solve the system of differential equations in Eqs. (\ref{eq:set_diff_eqs}) for varying initial conditions. Each solution then corresponds to a different trajectory plotted versus $n_w$ and $|a_-|$. All of these trajectories are seen to lie on the surface after the initial transition stage similarly to Fig. \ref{fig:unit_cellphc}. Next, we fit a polynomial with all the trajectories excluding the initial transition stage. Then the polynomial describes the surface $n_c$ in terms of $n_w$ and $\mathopen|a_-\mathclose|$. Similarly, we can approximate the surfaces for $\mathopen|a_+\mathclose|(\mathopen|a_-\mathclose|, n_w)$ and $\Delta\phi(\mathopen|a_-\mathclose|, n_w)$. We emphasize that in order to keep the original coordinate system of the variables, we exclude the part of the trajectory in the initial transition stage and fit a polynomial with the remaining parts of all the trajectories. Thus, we fit the polynomials once the state has reached the center manifold. Having obtained these surfaces, we can determine any state in the phase space within the periodic orbit once $n_w$ and $|a_-|$ are known without any need of solving the five-dimensional system of equations in Eq. (\ref{eq:set_diff_eqs}).

Moreover, in order to describe the dynamics on these surfaces, we need to solve a system of two differential equations describing the two degrees of freedom, $n_w$ and $|a_-|$. These equations are the components of the velocity vector in Eq. (\ref{eq:velocity2}):

\begin{subequations}\label{eq:simplified}
\begin{equation}
\frac{d|a_-(\tau)|}{d\tau}=V_2(|a_-(\tau)|,n_w(\tau)),
\end{equation}
\begin{equation}
\frac{dn_w(\tau)}{d\tau}=V_4(|a_-(\tau)|,n_w(\tau)),
\end{equation}
\end{subequations}

Once Eq. (\ref{eq:simplified}) is solved, the remaining degrees of freedom \{$|a_+|$, $\Delta\phi$, $n_c$\} can be determined using the polynomials.

\subsection{Instantaneous eigenvalues}

\begin{figure*}[t!]
    \subfloat[]{     \includegraphics[width=0.425\linewidth]{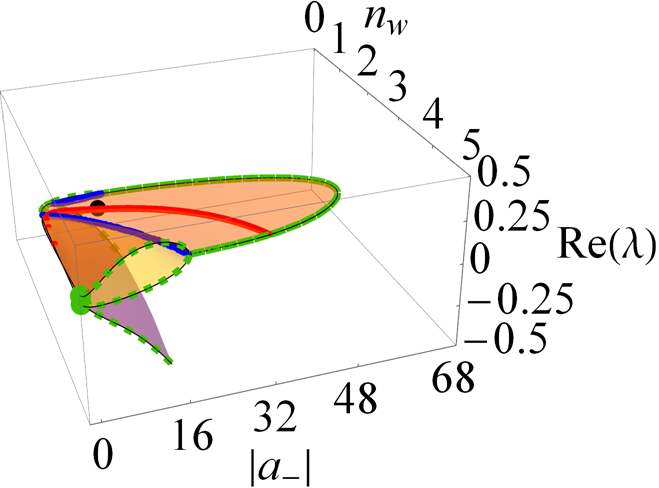}}
    \subfloat[]{     \includegraphics[width=0.425\linewidth]{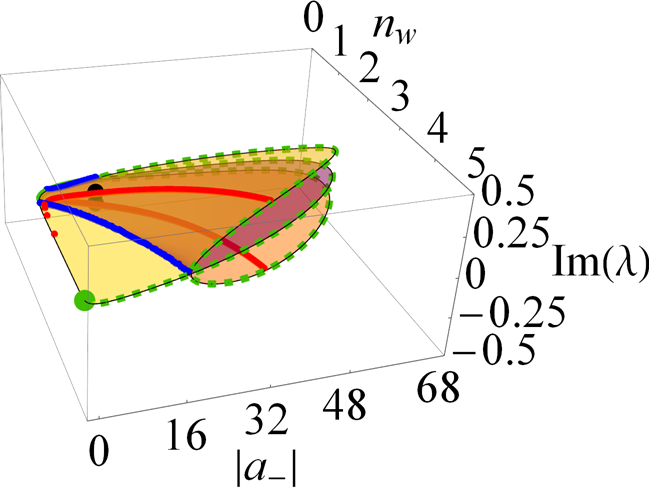}} \centering\\
    \subfloat[]{     \includegraphics[width=0.328\linewidth]{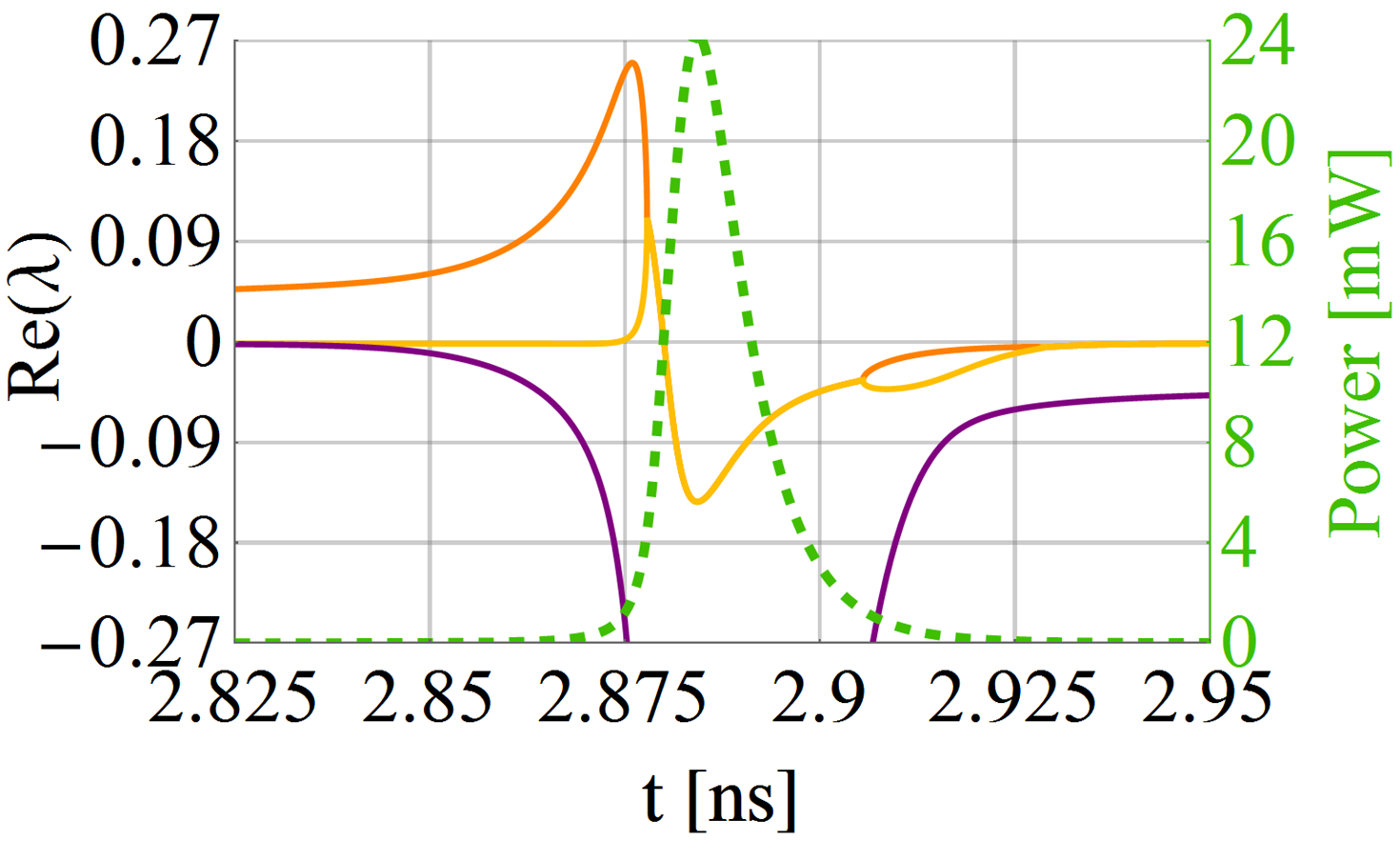}} \centering
    \subfloat[]{     \includegraphics[width=0.328\linewidth]{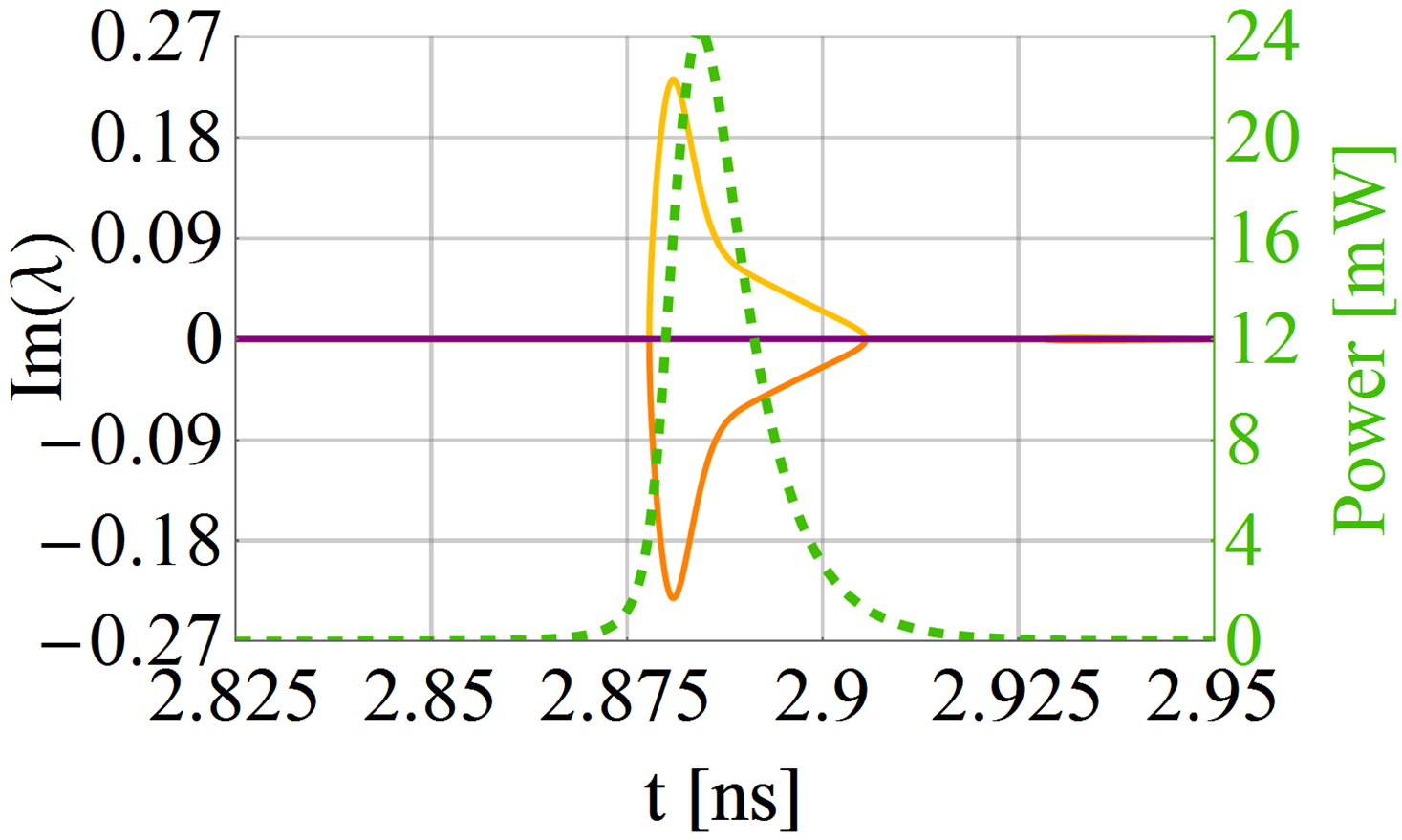}} \centering
        \subfloat[]{     \includegraphics[width=0.328\linewidth]{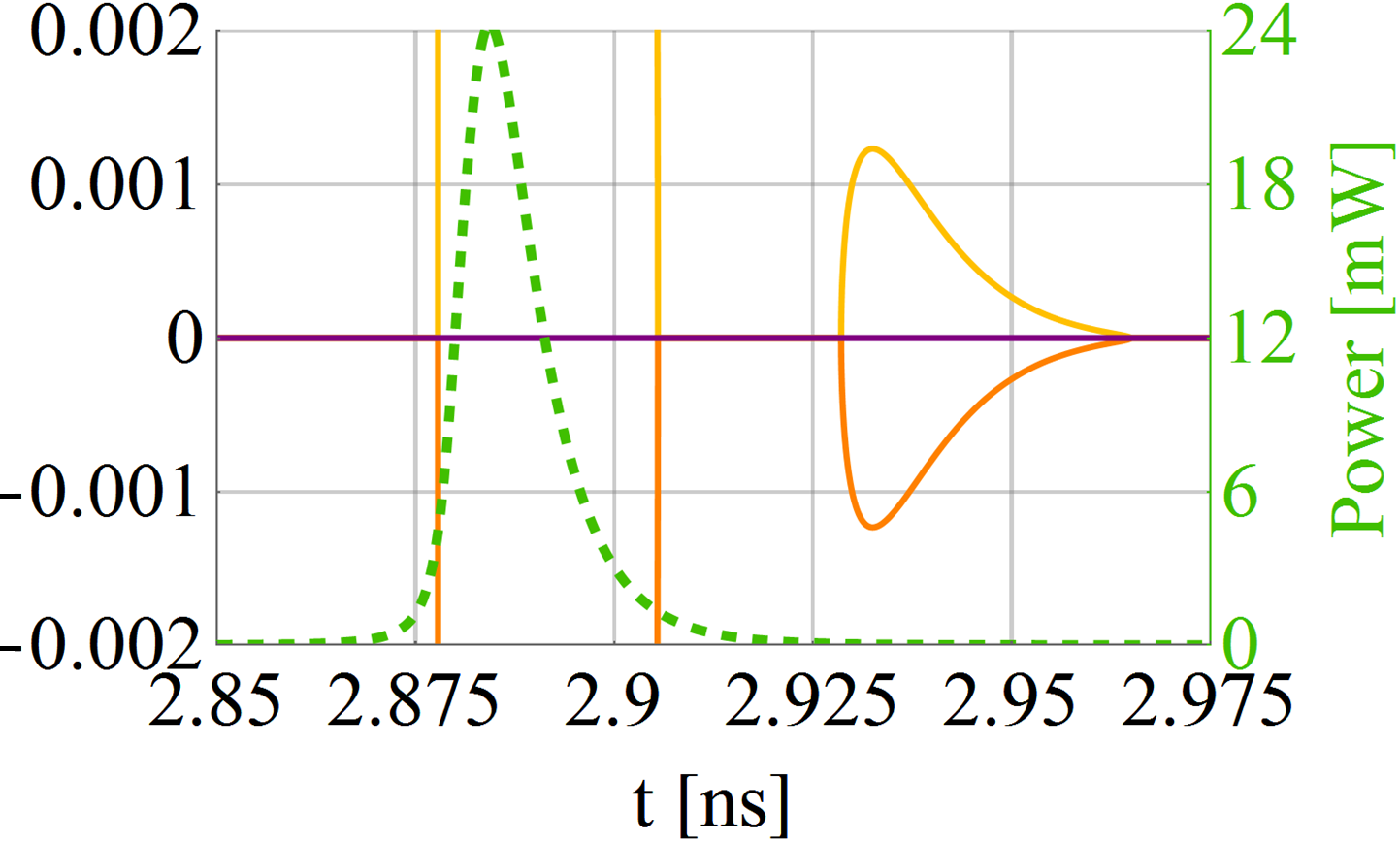}} \centering
  \caption{Real a) and imaginary b) parts of the three instantaneous eigenvalues, $\lambda$, of $\boldsymbol{A}$ with the largest real parts. The eigenvalues form a complex conjugate pair marked with orange and yellow, the remaining eigenvalue is purely real and marked with purple. The red line indicates the positions when $\mathrm{Re}(\lambda)=0$. The blue line indicates the contours of exceptional points. The black dot indicates the eigenvalues at the steady-state. The green dot indicates initial eigenvalues of the trajectory, marked with the dashed green line, plotted in c), d). c), d) show real and imaginary parts of the instantaneous eigenvalues with the largest real parts along the green trajectory in a), b). The pulse-power in the straight-port is marked with dashed green line (on the right axes). e) shows the imaginary parts of the eigenvalues in the vicinity of the second pair of the exceptional points. }
  \label{fig:instant_eig}
\end{figure*}

We compute the instantaneous eigenvalues of the matrix $\boldsymbol{A}$ in Eq. (\ref{eq:acceleration}) for the state vectors $\vec{\psi}$, Eq. (\ref{eq:position2}) over the whole surfaces. In Fig. \ref{fig:unit_cellphc}, we have seen that we can define three surfaces for $|a_+|$, $\Delta\phi$, $n_c$ plotted versus $n_w$ and $|a_-|$. After the initial transition stage, the state always lies on these surfaces. Thus, all the dynamics are confined to these surfaces. Then, each point of these surfaces can be substituted into the matrix $\boldsymbol{A}$ and the instantaneous eigenvalues of the state at this position are obtained. Figure \ref{fig:instant_eig} shows the real and imaginary parts of the three instantaneous eigenvalues with the largest real parts over the whole surface as well as along the trajectory marked with the green dashed line. The two remaining eigenvalues have significantly smaller real parts, and thus are not included in Fig. \ref{fig:instant_eig} as the contribution from the corresponding eigenvectors decays rapidly.

Figure \ref{fig:instant_eig}a) shows that the pair of eigenvalues marked with orange and yellow has considerably larger real parts than the third eigenvalue (purple) over the major part of the surface. The third eigenvalue is only comparable to the other two eigenvalues along the line $|a_-|=0$, but it is still smaller and never becomes positive. The negative real parts of the purely real third eigenvalue (purple) and the remaining complex conjugate pair of eigenvalues (not shown) signifies that the contribution of the corresponding eigenvectors in a reconstruction of the solution decays very quickly. This is what is observed in Fig. \ref{fig:unit_cellphc} in the initial transition stage. Afterwards, once the state is on the surface, the contribution from the three corresponding eigenvectors is negligible and the state description is dominated by the eigenvectors corresponding to the two eigenvalues with the largest real part (orange and yellow).

The real parts of the pair of eigenvalues marked with orange and yellow are seen to dominate for large values of $|a_-|$; this is where the pulse is released. In Fig. \ref{fig:instant_eig}c),d,e)) we show the instantaneous eigenvalues in the vicinity of the pulse along the green trajectory in a), b) when the state has already reached a limit cycle. On the right axis we plot the pulse power in the straight-port defined as:

\begin{equation}
P_+(t)=2\epsilon_0n_{ref}c_0|A_+(t)+PA_-(t)|^2,
\end{equation}

\noindent where $c_0$ is the speed of light and $\epsilon_0$ is vacuum permittivity.

Figure \ref{fig:instant_eig} shows that when the state moves along the $n_w$ axis (just after the previous pulse has been released and before a new one) the three eigenvalues with the largest real parts are purely real. As the limiting value of $n_w$ is reached, Fig. \ref{fig:instant_eig}a), one of the eigenvalues (orange) starts to rapidly increase, Fig. \ref{fig:instant_eig}c), while the third eigenvalue (purple) drops rapidly. Just as the pulse is released, the second eigenvalue (yellow) rapidly increases and collapses with the first one (yellow) at the exceptional point. Therefore, it is found that, as the pulse grows, the pair of eigenvalues transitions from being purely real to being complex conjugate when crossing the exceptional point. As the pulse power decreases, the complex conjugate pair of eigenvalues coalesce at the second exceptional point and transitions back to the pair of purely real eigenvalues. Thus, most of the pulse is observed to be bounded by the two instantaneous exceptional points with positive/negative real part of the eigenvalue at the beginning/end of the pulse, respectively. Interestingly, two more exceptional points are observed as the pulse is decaying, see Fig. \ref{fig:instant_eig}e).

Within one period the state traverses a loop in the phase space of the model. We have seen that four exceptional points are crossed within a single loop when the laser state is a periodic orbit. When the exceptional point is approached, the eigenvectors exhibit a characteristic phase jump and are phase shifted relative to each other by $\pm i$ \cite{gunther_projective_2007,heiss_repulsion_2000}. Therefore, during an evolution along any trajectory in the diminishing vicinity of an exceptional point, eigenvectors will acquire a phase shift of $\pm i$ \cite{keck_unfolding_2003,rotter_non-hermitian_2009,muller_exceptional_2008}. In \cite{menke_state_2016}, it has been shown that this effect is preserved as long as the exceptional point is inside the loop or crossed by it. Therefore, it is only a four-fold loop around an exceptional point or a single loop around four exceptional points that will restore an original scenario for the eigenvectors concerned \cite{heiss_collectivity_1998,heiss_phases_1999,dembowski_experimental_2001, heiss_exceptional_2004, heiss_physics_2012}. Since the laser is operating in the periodic orbit in our case, in order to remain periodic it has to cross four exceptional points within one period in phase space.

\subsection{Reconstruction of the solution}

At most two out of five instantaneous eigenvalues have positive real parts. Thus, after the initial transition stage, the eigenvectors which correspond to the two dominating eigenvalues can be used to reconstruct the solution of Eq. (\ref{eq:acceleration}) as follows:

\begin{equation}\label{eq:eig_expansion}
\frac{d\vec{\psi}}{d\tau}=c_1(\tau)\vec{v_1}(\tau)+c_2(\tau)\vec{v_2}(\tau)
\end{equation}

\noindent where $\vec{v}_{1,2}$ are the instantaneous right eigenvectors, and $c_{1,2}$ are the amplitudes of the corresponding eigenvectors. These amplitudes can be reconstructed from a solution $d\vec{\psi}(\tau)/d\tau$ using the left eigenvectors as $c_{1,2}=\vec{w}_{1,2}^T(\tau)d\vec{\psi}(\tau)/d\tau$. In the following we show that the two eigenvectors can be used to approximate the two tangential vectors to the surface pointing along the $|a_-|$, $n_w$ coordinate lines. This confirms that the solution can be approximately expanded in the two eigenvectors.

The tangential vector to the surface $z=f(x,y)$ along the parametric curve $\vec{r}(t)=\{ x(t), y(t), z(t)\}$ on this surface is expressed as $\vec{r}'(t)=\{ x'(t), y'(t), z'(t)\}$, where $z'(t)=\nabla f \cdot\vec{u}$, $\vec{u}=\{ x'(t), y'(t)\}$. In our case, the tangential vectors to the surfaces, which approximate the components of the state vector $\vec{\psi}$, Eq. (\ref{eq:position2}), are expressed as:

\begin{figure*}[t!]
 \centering
     \subfloat[]{\includegraphics[width=0.3325\linewidth]{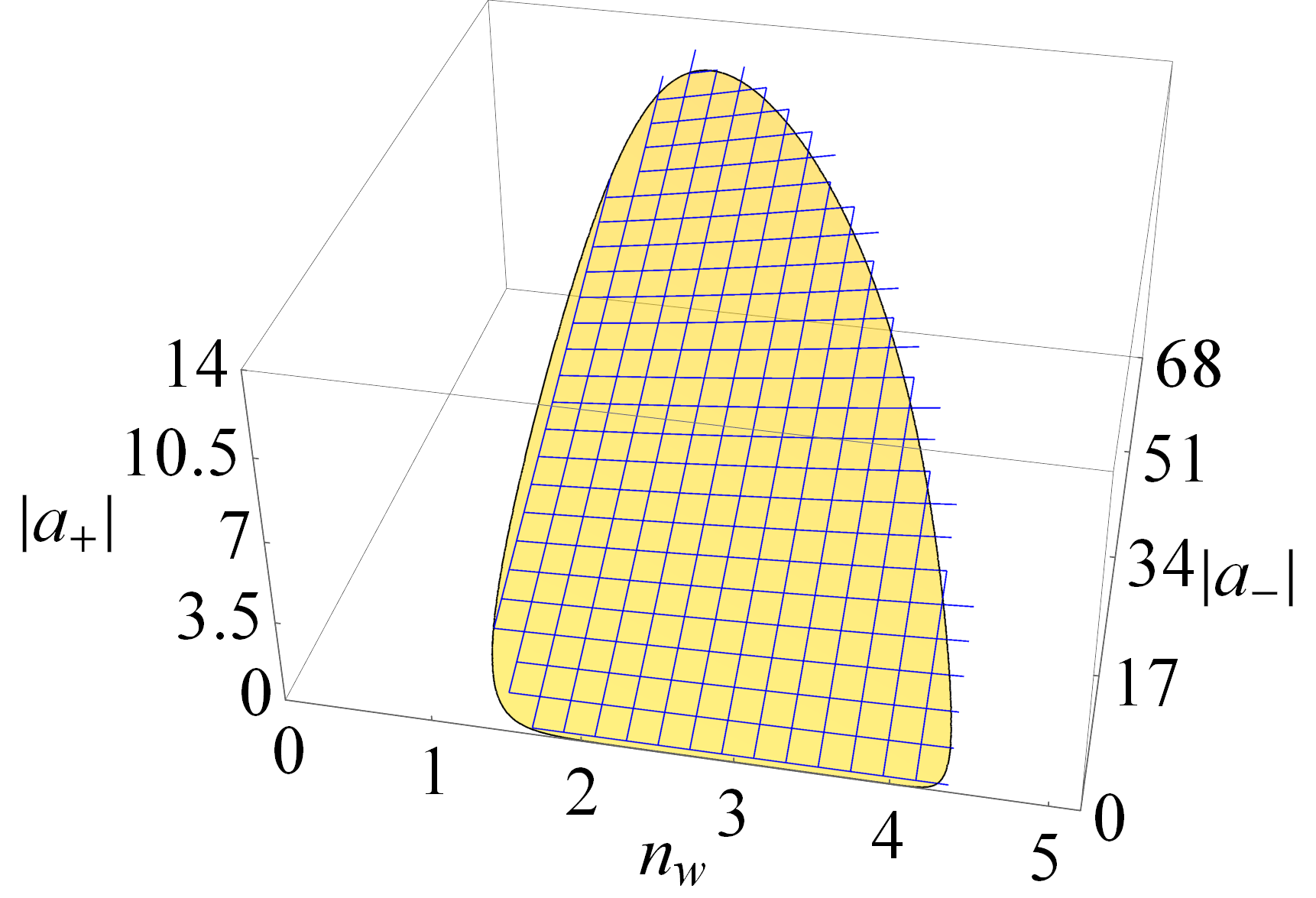}}\hfill
     \subfloat[]{\includegraphics[width=0.3225\linewidth]{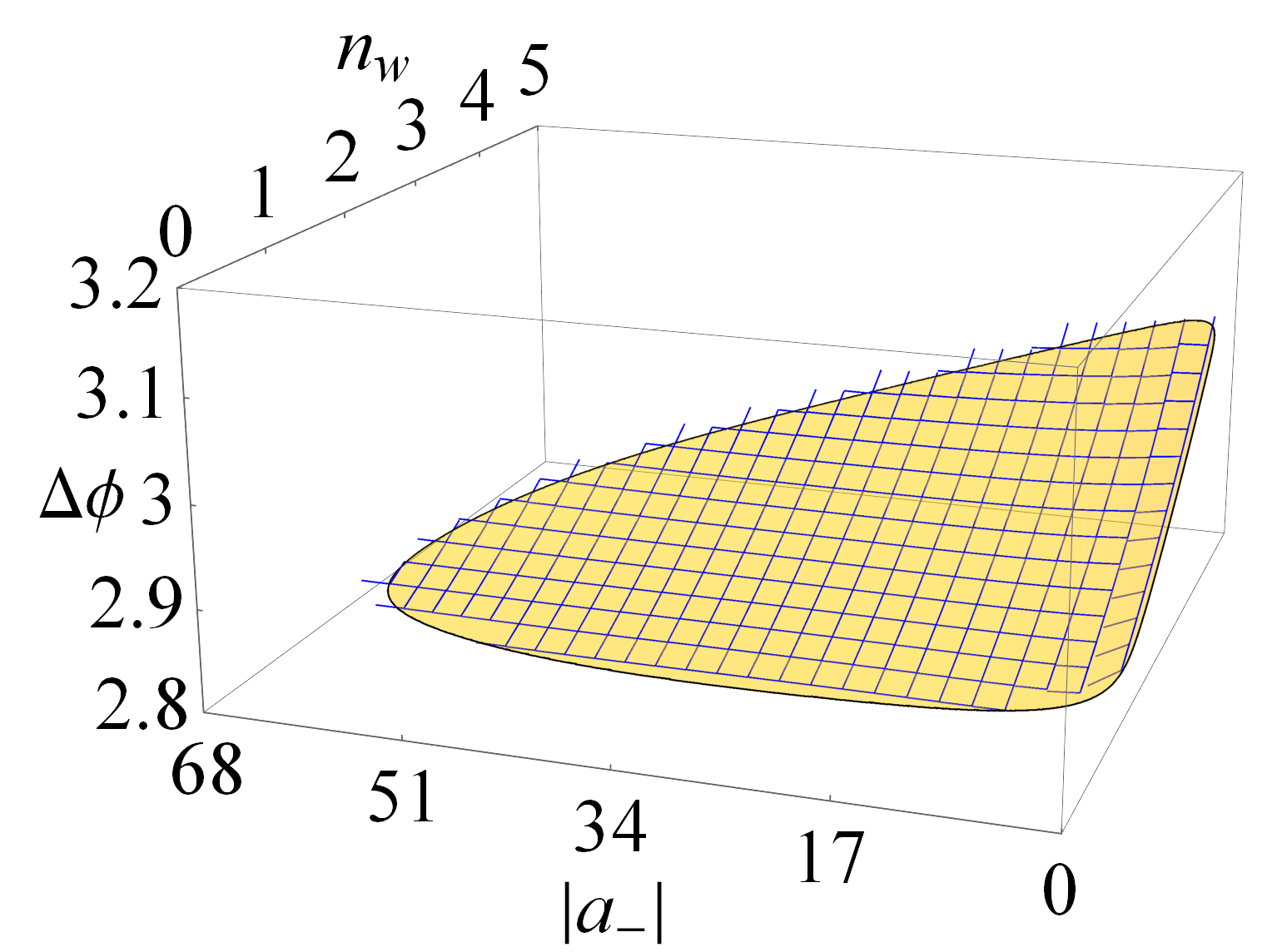}}\hfill
     \subfloat[]{\includegraphics[width=0.3325\linewidth]{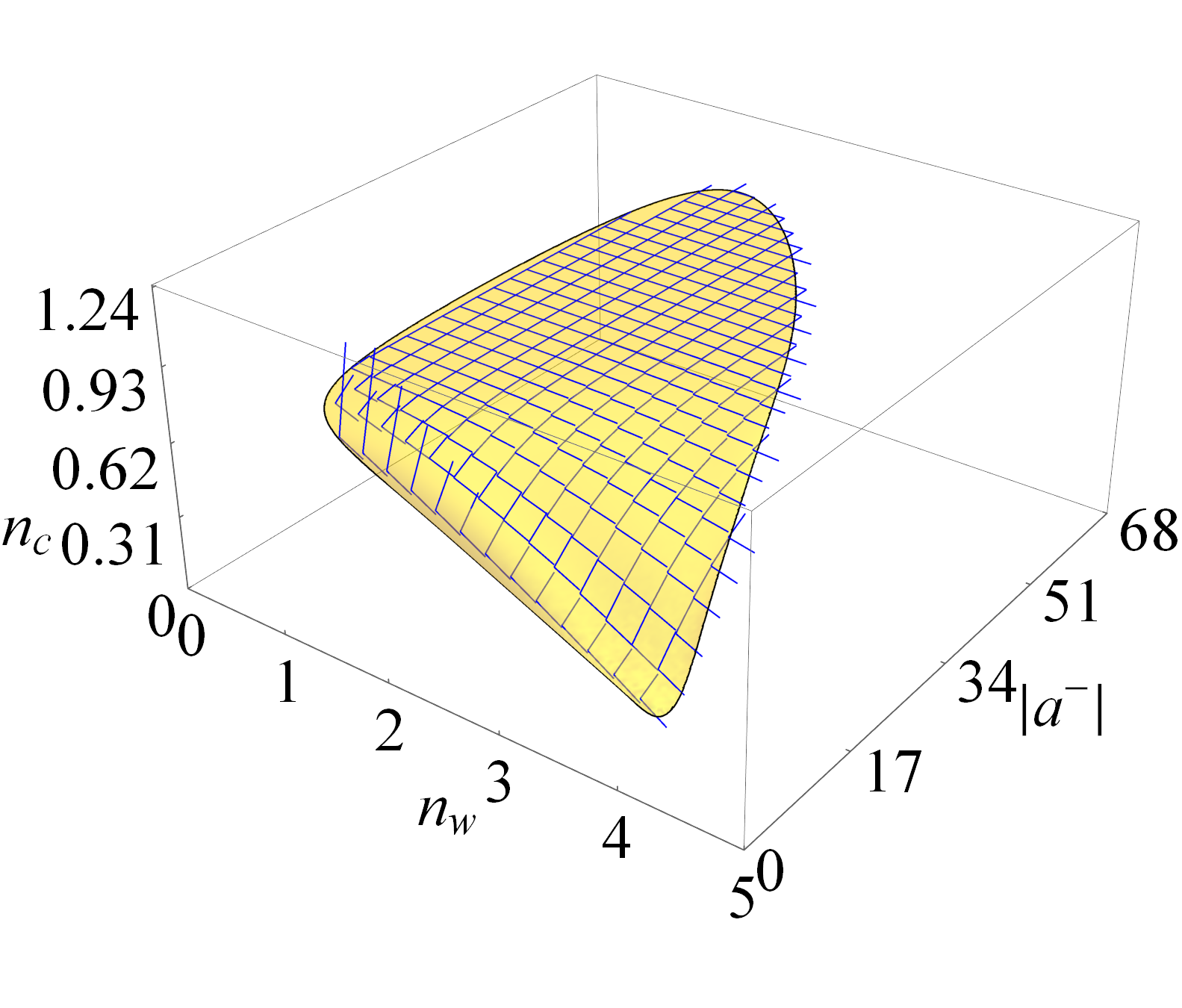}}
  \caption{The vectors $\vec{v}_{1,2}'$ plotted on the three surfaces, a) $|a_+|$, b) $\Delta\phi$, c) $n_c$. The vectors $\vec{v}_{1,2}'$ result from the linear combination of the eigenvectors corresponding to the top two eigenvalues and enforcing them to point along the $n_w$, $|a_-|$ coordinate lines. These vectors are found to approximate the tangential vectors to the surface along the $n_w$, $|a_-|$ coordinate lines.}
  \label{fig:tangent_eig}
\end{figure*}

\begin{equation}\label{eq:tang_vec1}
\vec{r}'(\tau)=\bigg\{ \frac{d|a_-|}{d\tau}, \frac{dn_w}{d\tau}, \frac{d\vec{\psi}}{d\tau}\bigg\},
\end{equation}

\noindent where

\begin{equation}\label{eq:tang_vec2}
\frac{d\vec{\psi}}{d\tau}=\frac{\partial\vec{\psi}}{\partial|a_-|}\frac{d|a_-|}{d\tau}+\frac{\partial\vec{\psi}}{\partial n_w}\frac{dn_w}{d\tau}.
\end{equation}

The tangential vectors $\vec{r}'(\tau)$ along the parameterized trajectory can be decomposed into a linear combination of the tangential vectors to the surface along its coordinates $|a_-|$, $n_w$:

\begin{equation}\label{tange_vec3}
\vec{r}'_1=\bigg\{ 1, 0, \frac{\partial\vec{\psi}}{\partial|a_-|}\bigg\}, \hspace{0.65cm} \vec{r}'_2=\bigg\{ 0, 1, \frac{\partial\vec{\psi}}{\partial n_w}\bigg\},
\end{equation}

It is observed that the tangential vectors to the surface are composed of the components of the velocity vector, Eq. (\ref{eq:tang_vec1}), and the velocity vector can be expanded into the two eigenvectors, see Eq. (\ref{eq:eig_expansion}). Since the five-dimensional (5D) matrix $\boldsymbol{A}$ is real, the top two eigenvalues ($\lambda_1$ and $\lambda_2$) and eigenvectors ($\vec{v_1}$ and $\vec{v_2}$) are either real or form a complex conjugate pair. Then, we change these eigenvectors to point along the original coordinate lines, $|a_-|$, $n_w$, as follows:

\begin{equation}\label{lincomb_eig}
\begin{bmatrix}
  \vec{v}_1' \\
  \vec{v}_2'
\end{bmatrix}
=
\begin{bmatrix}
  v_{12} & v_{14} \\
  v_{22} & v_{24}
\end{bmatrix}^{-1}
\begin{bmatrix}
  v_{12} & v_{14} & v_{11} & v_{12} & v_{13} & v_{14} & v_{15} \\
  v_{22} & v_{24}  & v_{21} & v_{22} & v_{23} & v_{24} & v_{25}
\end{bmatrix}
\end{equation}

 Then, the two vectors $\vec{v}_1'$ and $\vec{v}_2'$ are determined at positions of the state vector approximated by the polynomials, see Eq. (\ref{eq:position2}), and separated by equidistant steps. The vectors are purely real and are plotted over the whole surfaces $|a_+|$, $\Delta\phi$, $n_c$, see Fig. \ref{fig:tangent_eig}. Subsequently, these vectors are scaled by the distance between the steps in the state vector along each direction in order to avoid an overlap and create a square grid pattern. If these vectors create an ideal square grid then they can perfectly reconstruct the tangential vectors in Eq. (\ref{tange_vec3}). A small discrepancy is only found in Fig. \ref{fig:tangent_eig}(c) for small values of $|a_-|$, which can be explained by the third eigenvalue becoming comparable to the dominating pair of eigenvalues at these points, see Fig. \ref{fig:instant_eig}. However, the two vectors $\vec{v}_1'$ and $\vec{v}_2'$ are found to approximate the tangential vectors over the whole surface as observed in Figure \ref{fig:tangent_eig}. Thus, the two degree-of-freedom picture is justified over the whole surface and shown to precisely reconstruct $d\vec{\psi}(\tau)/d\tau$. Therefore, the system of five nonlinear differential equations can be reduced to only two differential equations after the initial transition stage. The other three dimensions are functions of $n_w$ and $|a_-|$ and are presently approximated by polynomials. We note that the instantaneous eigenvalues/eigenvectors are not needed to reduce dimensionality of the system, but they provide an additional insight into the solution. Furthermore, the fact that the two instantaneous eigenvectors approximate the tangential vectors to the surfaces proves that the system dimensionality can be reduced to two.

Moreover, we note that although a 2D model can be used to describe the laser dynamics after the initial transition stage, there exists a parameter region in which even a 1D model is sufficient to replace the original 5D model after the initial transition stage. One may observe in Fig. \ref{fig:steady_state_eig} that for a large negative detuning $\Delta\omega_c$, the steady-state eigenvalues undergo transition from a complex conjugate pair of eigenvalues to two purely real eigenvalues. Then, one of the eigenvalues decreases rapidly, and the other one approaches zero. Therefore, for detunings $-2.05\gamma_T<\Delta\omega_c<-1.72\gamma_T$, there is a single steady-state eigenvalue that dominates and thus, the velocity vector can be described by a single eigenvector, see Eq. (\ref{eq:eig_expansion}). In this case, the laser dynamics can be described by a single differential equation after the transition stage in which the contribution from the other four eigenvalues rapidly decays. For detunings $\Delta\omega_c<-2.05\gamma_T$, the lasing mode ceases to exist \cite{mork_chaos_1992,mork_photonic_2014,rasmussen_theory_2017}. Thus, as the detuning $\Delta\omega_c$ increases, the steady-state eigenvalues transition from being purely real to a complex conjugate pair and the system evolves from a 1D to a 2D system.

\section{\label{sec:Hopf}Origin of the laser instability}

\subsection{Detection of periodic orbits}

In what follows, we use the simplified 2D model, Eq. \ref{eq:simplified}, to explain the origin of the laser instability that may be observed even when all real parts of the steady-state eigenvalues are negative.

\begin{figure}
      \subfloat[]{     \includegraphics[width=0.89\linewidth]{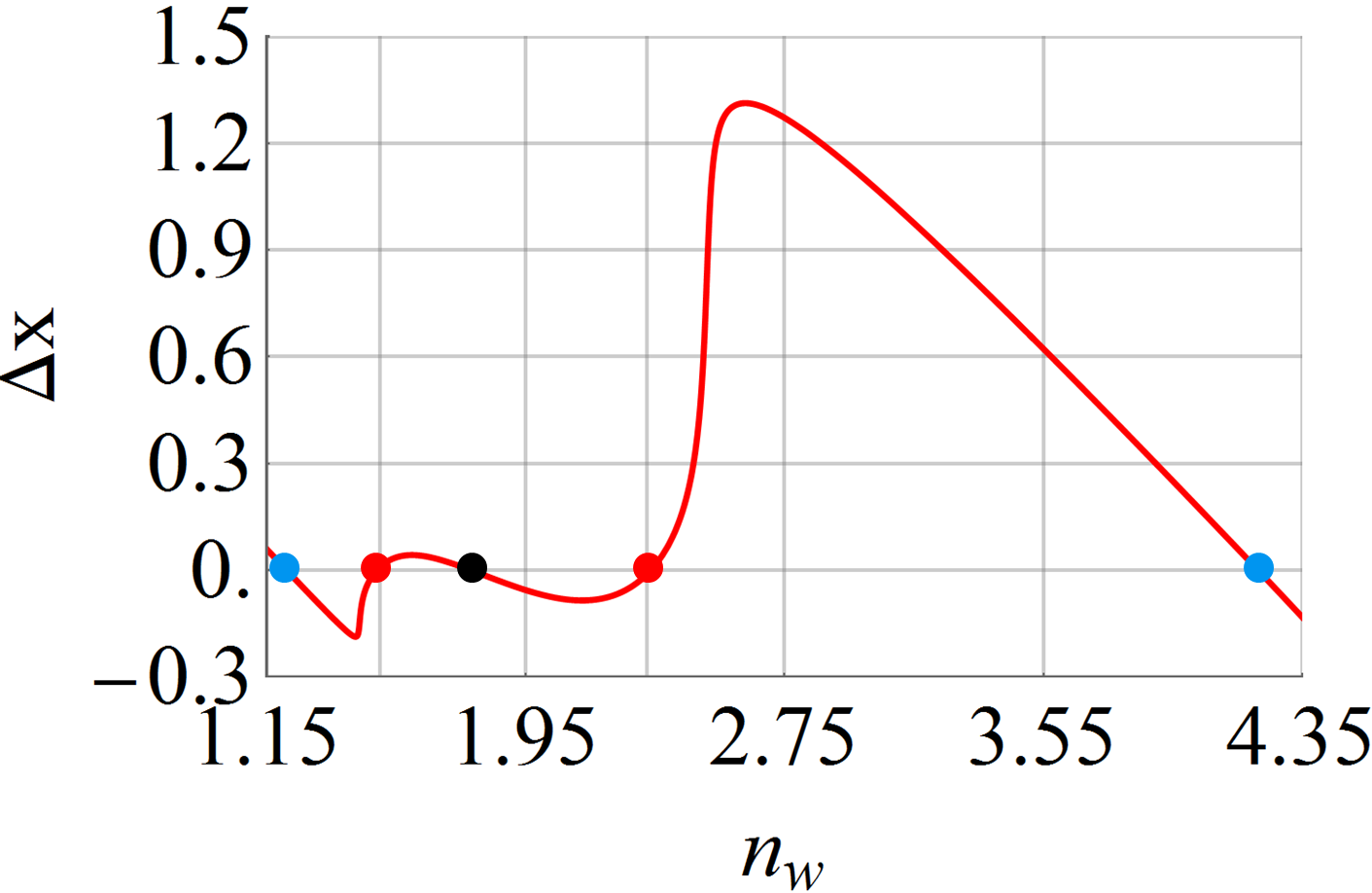}} \centering\\
      \subfloat[]{     \includegraphics[width=0.89\linewidth]{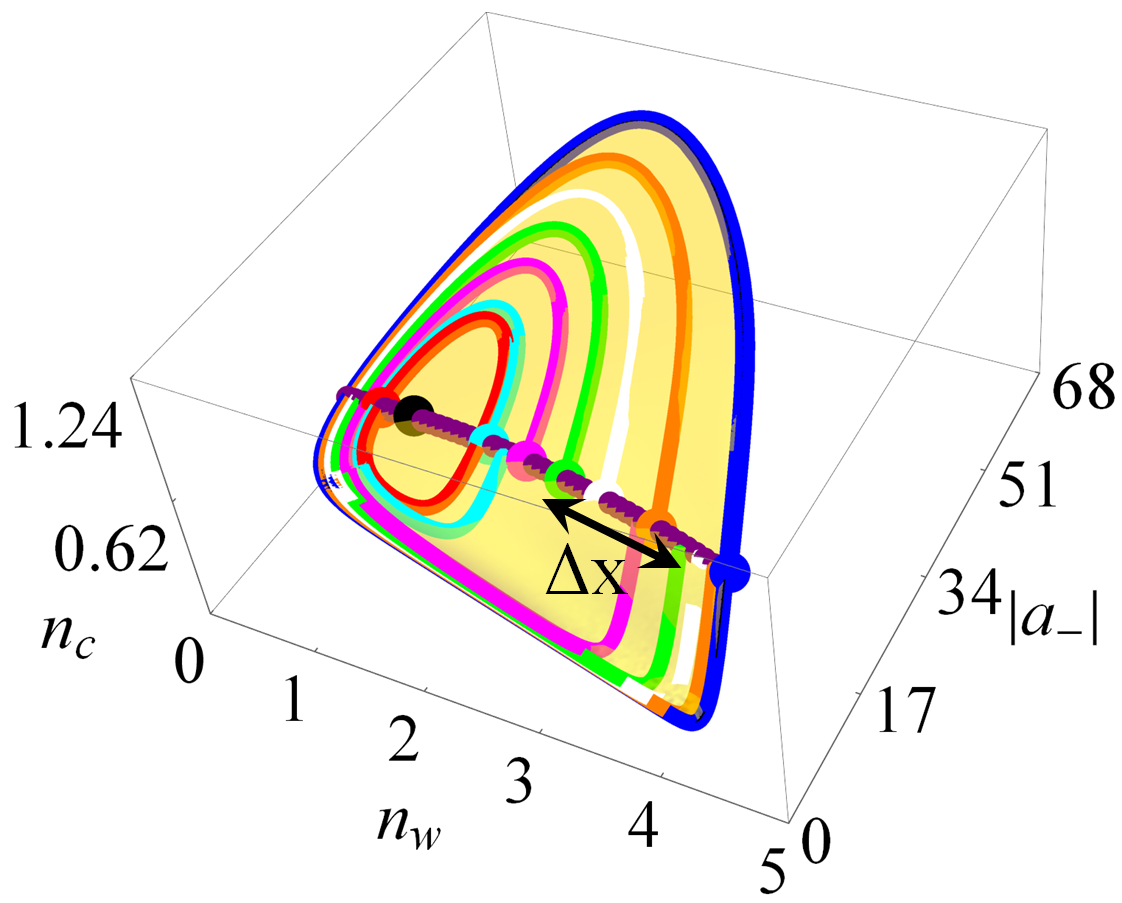}} \centering
  \caption{a) Numerically evaluated shift $\Delta x=n_w^{\mathrm{initial}}-n_w^{\mathrm{cycle}}$ in phase space after one cycle of the trajectory. The single black dot indicates the steady state, the pairs of red and blue dots indicate two periodic orbits. The initial value for $|a_-|$ is fixed and equal to the steady-state value, while the initial value for $n_w$ is varied. The initial states are marked on the surface along the purple line in b). b) Figure showing a single cycle of several trajectories (marked with different colours) initiated at different initial states along the purple line on the curved surface $n_c(|a_-|,n_w)$. The initial states are marked with dots. The definition of $\Delta x$ is also indicated. }
  \label{fig:periodic_orbits}
\end{figure}

At first, the phase space of the Fano laser is scanned in search for periodic orbits. We choose our initial conditions as follows: $1)$ $|a_-|^{\mathrm{initial}}$ is set to the steady-state value and $2)$ $n_w^{\mathrm{initial}}$ is varied over the whole phase space along the purple line as shown in Fig. \ref{fig:periodic_orbits}b). For each initial condition, we then compute the trajectory by solving Eq. (\ref{eq:simplified}) up to the point when $|a_-|(T)=|a_-|^{\mathrm{initial}}$, where $T$ is the time corresponding to one cycle. Some of these trajectories are shown in Fig. \ref{fig:periodic_orbits}(b) in different colours. Subsequently, we evaluate the shift $\Delta x=n_w^{\mathrm{initial}}-n_w^{\mathrm{cycle}}$ in the state vector after the time $T$. If the shift between the initial state and the state after one cycle is zero then we are at a periodic orbit or steady-state. On the other hand if it is non-zero it means that the state is approaching or departing from the steady-state/periodic orbit.

Figure \ref{fig:periodic_orbits}a) shows the numerically evaluated shift $\Delta x$ in $n_w$ after one cycle of the trajectory. It is seen that there are five crossings with zero. These crossings are marked with blue, red and black dots. The single black dot indicates the steady-state, while the pairs of blue and red dots indicate periodic orbits. The outer periodic orbit, marked with a pair of blue dots, has been observed before and is known to be stable for a pair of complex conjugate steady-state eigenvalues with a positive real part \cite{rasmussen_theory_2017}. Here, it is seen that for a strong enough perturbation of the initial conditions from the steady-state value, the state can still reach the outer periodic orbit despite all the steady-state eigenvalues having a negative real part and thus the steady-state being stable and attracting the state. Furthermore, we find an additional periodic orbit marked with a pair of red dots in Fig. \ref{fig:periodic_orbits}a). The newly found periodic orbit separates the steady-state and the outer periodic orbit.

\subsection{Stability of the orbits}

We now prove the stability of the newly found orbit using the simplified 2D model. This is done by calculating the Floquet multipliers, $\lambda_f$, which tell us how the solution behaves in the vicinity of the periodic orbit, i.e., whether it diverges/converges from/towards the orbit \cite{iooss_elementary_1997,glendinning_stability_1994}. In order to compute the Floquet multipliers, we first obtain the fundamental solution matrix, $\boldsymbol{\Phi}(\tau)$, which can be determined using $d\boldsymbol{\Phi}(\tau)/d\tau=\boldsymbol{A}(\tau)\boldsymbol{\Phi}(\tau)$ and satisfies $d\vec{\psi}/d\tau=\boldsymbol{\Phi}(\tau)d\vec{\psi}/d\tau\bigl|_{\tau=0}$ with $\boldsymbol{\Phi}(0)=\boldsymbol{I}$. The Floquet multipliers are the eigenvalues of $\boldsymbol{\Phi}(\tau)$ evaluated at $\tau=T$, where $T$ is the period of the orbit.

If the Floquet multipliers are within the unit circle in the complex plane, the orbit is stable, otherwise it is unstable. The Floquet multipliers for the outer periodic orbit are $\lambda_{f1}=0.04$ and $\lambda_{f2}=1$ confirming its stability. On the other hand, the Floquet multipliers of the newly found orbit are $\lambda_{f1}=2.31$ and $\lambda_{f2}=1$, proving that this orbit is unstable. We note that for a periodic orbit there is always one of the Floquet multipliers for which $\lambda_f=1$ and the corresponding eigenvector is tangential to the periodic orbit. This neutral stability accounts for the possibility of drift along the periodic orbit \cite{glendinning_stability_1994}.

Furthermore, we study the stability of the newly found orbit with variation of the detuning, $\Delta\omega_c$. It is observed that the Floquet multiplier crosses the unit circle along the real axis in the complex plane. This indicates an exchange of instability \cite{seydel_practical_2010}. Indeed, as $\Delta\omega_c$ decreases from $\Delta\omega_c=0.52\gamma_T$, see Fig. \ref{fig:steady_state_eig}, the newly found unstable periodic orbit increases in size. Eventually, it collapses with the stable periodic orbit. Both orbits disappear due to a fold bifurcation \cite{kuznetsov_elements_2004} and only the stable (time-independent) steady-state remains present in the phase space. On the other hand, when $\Delta\omega_c$ increases, the new unstable orbit decreases in size. Eventually it collapses with the stable steady-state resulting in the steady-state becoming unstable. This happens in the vicinity of $\Delta\omega_c=1.52\gamma_T$, which is the critical bifurcation point and as $\Delta\omega_c$ is increased further, the real part of the steady-state eigenvalues becomes positive. Since the cycle is present before the bifurcation point, i.e. before the real parts of the steady-state eigenvalues become positive, the bifurcation at this point is called a subcritical Hopf bifurcation.

\begin{figure}
\includegraphics[width=0.999\linewidth]{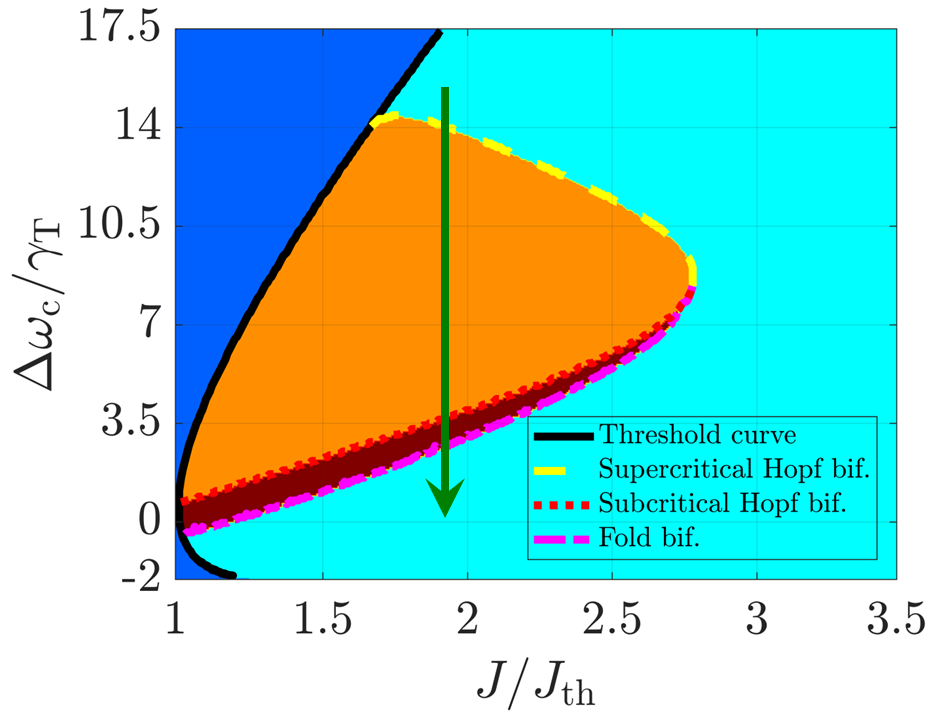}
\caption{Phase diagram of the Fano laser as a function of pump current $J/J_{\mathrm{thr}}$ and cavity detuning $\Delta\omega_{\mathrm{c}}/\gamma_{\mathrm{T}}$. Blue indicates solutions below threshold, cyan marks a stable continuous-wave solution. Orange indicates the presence of a stable limit cycle and an unstable steady-state, while dark red indicates two orbits being present, stable and unstable, as well as a stable steady-state. The laser threshold curve is shown in black. The dashed yellow line and dotted red line mark supercritical and subcritical Hopf bifurcations respectively. The dash-dotted magenta line indicates a fold bifurcation.}
\label{fig:phase_diag}
\end{figure}

Both bifurcations are marked in the phase diagram in Fig. \ref{fig:phase_diag}. It shows that as $\Delta\omega_c$ is decreased from large values along the dark green arrow, at first the system undergoes a supercritical Hopf bifurcation at the dashed yellow line. There, the real part of the steady-state eigenvalues crosses zero and becomes positive resulting in the steady-state point becoming unstable and a stable limit cycle being present after the bifurcation point. As we decrease $\Delta\omega_c$ further, the system undergoes a subcritical Hopf bifurcation at the dotted red line. Here, the steady-state point becomes stable again, while the stable periodic orbit coexists with an unstable periodic orbit. Eventually, as $\Delta\omega_c$ is further decreased, the unstable periodic orbit collides with the stable one, and both orbits disappear through a fold bifurcation leaving the stable steady-state point as the only solution \cite{kuznetsov_elements_2004, seydel_practical_2010}. Analogous behaviour is observed for lower pump currents including $J=1.2J_{\mathrm{thr}}$, however in this case the laser is below threshold for large detuning $\Delta\omega_c$.

Occurrence of both Hopf bifurcations, supercritical and subcritical, is a signature of a Bautin bifurcation, also known as generalized Hopf bifurcation \cite{govaerts_numerical_2000,kuznetsov_elements_2004}. A Bautin bifurcation is characterized by the presence of two orbits and an equilibrium point (steady-state) in phase space. We note that a Bautin bifurcation cannot be detected by merely monitoring the eigenvalues \cite{govaerts_numerical_2000,kuznetsov_elements_2004}. Upon the external parameter variation, $\Delta\omega_c$, an inner orbit may collide with an outer orbit and annihilate or exchange stability with an equilibrium point, as it has been observed. We note that since each $\Delta\omega_c$ results in different solutions $(\omega_s, N_s)$ of the oscillation condition in Eq. (\ref{eq:osc_cond}), we adjust the polynomial approximation of the surface in each case.

The stability of the orbits can also be assessed based on Fig. \ref{fig:periodic_orbits}a). It is seen that if the model is initialized outside the orbit marked with the red dots, $n_w^{\mathrm{cycle}}$ will increase after one cycle compared to the initial value. Thus, the state is repelled away from the orbit. If the model is initialized inside the orbit,  $n_w^{\mathrm{cycle}}$ will keep increasing with each cycle confirming that the newly found orbit is unstable.

\section{Conclusion}
We demonstrate that after a fast initial transient the dynamics of the recently realized Fano laser \cite{yu_demonstration_2017} are confined to a 2D center manifold. The dimension of the center manifold follows the number of steady-state eigenvalues, the real parts of which cross zero. We show that there are two steady-state eigenvalues with real parts crossing zero, while the remaining three eigenvalues have negative real parts forming a stable manifold. The dynamics is attracting along the corresponding three directions and quickly tends to the curved surface, i.e. the center manifold, during the initial transition stage. Afterwards, the state vector is confined to the curved surface and can be solely described by two degrees of freedom. The surface geometry of the phase space can be approximated by the two eigenvectors of the linear stability matrix corresponding to the eigenvalues with the largest real parts. As the pulse develops, the instantaneous eigenvalues transition from a pair of purely real eigenvalues to a complex conjugate pair at the first exceptional point. The main part of the repeating pulse is bounded by two exceptional points with positive/negative real part of the eigenvalue at the beginning/end of the pulse, respectively. Moreover, the trajectory encounters four exceptional points during one period, ensuring that both, the eigenvalues and the eigenvectors, are periodic in $\tau$. Furthermore, we show that the 5D model used to describe the laser dynamics, after the initial transition stage, can be reduced to only 1D in part of the parameter space and evolves into a 2D model beyond the exceptional point of steady-state eigenvalues as the detuning $\Delta\omega_c$ increases. Moreover, we have used the simplified 2D model to associate the unknown source of laser instability with the newly found unstable periodic orbit, which arises due to a generalized Hopf (Bautin) bifurcation. These findings allow to better understand the laser dynamics and may lead to the design of new functionalities in nanolasers used for on-chip communications and sampling.

\begin{acknowledgments}
The authors would like to thank T. S. Rasmussen for helpful discussions on the implementation of the dynamic model. This work was supported by Villum Fonden via the Centre of Excellence NATEC (grant 8692) and Research Grants Council of Hong Kong through project C6013-18GF.
\end{acknowledgments}

%\appendix

%\section{Appendixes}

%\nocite{*}


\begin{thebibliography}{10}

\bibitem{miller_device_2009}
D.~A.~B. Miller.
\newblock Device {Requirements} for {Optical} {Interconnects} to {Silicon}
  {Chips}.
\newblock {\em Proceedings of the IEEE}, 97(7):1166--1185, July 2009.

\bibitem{akahane_high-q_2003-1}
Yoshihiro Akahane, Takashi Asano, Bong-Shik Song, and Susumu Noda.
\newblock High-\textit{{Q}} photonic nanocavity in a two-dimensional photonic
  crystal.
\newblock {\em Nature}, 425(6961):944--947, October 2003.

\bibitem{tran_directive_2009}
Nguyen-Vi-Quynh Tran, Sylvain Combri\'e, and Alfredo De~Rossi.
\newblock Directive emission from high-$q$ photonic crystal cavities through
  band folding.
\newblock {\em Phys. Rev. B}, 79:041101(R), Jan 2009.

\bibitem{matsuo_ultralow_2013}
S.~Matsuo, T.~Sato, K.~Takeda, A.~Shinya, K.~Nozaki, H.~Taniyama, M.~Notomi,
  K.~Hasebe, and T.~Kakitsuka.
\newblock Ultralow {Operating} {Energy} {Electrically} {Driven} {Photonic}
  {Crystal} {Lasers}.
\newblock {\em IEEE J. Sel. Top. Quant.}, 19(4):4900311--4900311, July 2013.

\bibitem{jang_sub-microwatt_2015}
Hoon Jang, Indra Karnadi, Putu Pramudita, Jung-Hwan Song, Ki~Soo~Kim, and
  Yong-Hee Lee.
\newblock Sub-{microWatt} threshold nanoisland lasers.
\newblock {\em Nat. Commun.}, 6(1), December 2015.

\bibitem{hamel_spontaneous_2015}
Philippe Hamel, Samir Haddadi, Fabrice Raineri, Paul Monnier, Gregoire
  Beaudoin, Isabelle Sagnes, Ariel Levenson, and Alejandro~M. Yacomotti.
\newblock Spontaneous mirror-symmetry breaking in coupled photonic-crystal
  nanolasers.
\newblock {\em Nat. Photonics}, 9(5):311--315, May 2015.

\bibitem{mork_photonic_2014}
J.~Mork, Y.~Chen, and M.~Heuck.
\newblock Photonic crystal fano laser: Terahertz modulation and ultrashort
  pulse generation.
\newblock {\em Phys. Rev. Lett.}, 113:163901, Oct 2014.

\bibitem{fano_effects_1961}
U.~Fano.
\newblock Effects of {Configuration} {Interaction} on {Intensities} and {Phase}
  {Shifts}.
\newblock {\em Physical Review}, 124(6):1866--1878, December 1961.

\bibitem{limonov_fano_2017}
Mikhail~F. Limonov, Mikhail~V. Rybin, Alexander~N. Poddubny, and Yuri~S.
  Kivshar.
\newblock Fano resonances in photonics.
\newblock {\em Nat. Photonics}, 11(9):543--554, September 2017.

\bibitem{yu_demonstration_2017}
Yi~Yu, Weiqi Xue, Elizaveta Semenova, Kresten Yvind, and Jesper Mørk.
\newblock Demonstration of a self-pulsing photonic crystal {Fano} laser.
\newblock {\em Nat. Photonics}, 11(2):81--84, February 2017.

\bibitem{rasmussen_theory_2017}
Thorsten~S. Rasmussen, Yi~Yu, and Jesper Mørk.
\newblock Theory of {Self}-pulsing in {Photonic} {Crystal} {Fano} {Lasers}
\newblock {\em Laser Photonics Rev.}, 11(5):1700089, September 2017.

\bibitem{rasmussen_modes_2018}
Thorsten~S. Rasmussen, Yi~Yu, and Jesper Mørk.
\newblock Modes, stability, and small-signal response of photonic crystal
  {Fano} lasers.
\newblock {\em Opt. Express}, 26(13):16365, June 2018.

\bibitem{mork_chaos_1992}
J.~Mørk, B.~Tromborg, and J.~Mark.
\newblock Chaos in semiconductor lasers with optical feedback: theory and
  experiment.
\newblock {\em IEEE Journal of Quantum Electronics}, 28(1):93--108, January
  1992.

\bibitem{krauskopf_bifurcation_2000}
Bernd Krauskopf.
\newblock Bifurcation analysis of laser systems.
\newblock In {\em {AIP} {Conference} {Proceedings}}, volume 548, pages 1--30,
  Texel, (The Netherlands), 2000. AIP.

\bibitem{wieczorek_dynamical_2005}
S.~Wieczorek, B.~Krauskopf, T.B. Simpson, and D.~Lenstra.
\newblock The dynamical complexity of optically injected semiconductor lasers.
\newblock {\em Physics Reports}, 416(1-2):1--128, September 2005.

\bibitem{erzgraber_bifurcation_2007}
Hartmut Erzgräber, Bernd Krauskopf, and Daan Lenstra.
\newblock Bifurcation {Analysis} of a {Semiconductor} {Laser} with {Filtered}
  {Optical} {Feedback}.
\newblock {\em SIAM Journal on Applied Dynamical Systems}, 6(1):1--28, January
  2007.

\bibitem{govaerts_numerical_2000}
W.~Govaerts, Yu.~A. Kuznetsov, and B.~Sijnave.
\newblock Numerical {Methods} for the {Generalized} {Hopf} {Bifurcation}.
\newblock {\em SIAM Journal on Numerical Analysis}, 38(1):329--346, January
  2000.

\bibitem{kuznetsov_elements_2004}
Yuri Kuznetsov.
\newblock {\em Elements of {Applied} {Bifurcation} {Theory}}.
\newblock Springer, New York, 3rd edition edition, June 2004.

\bibitem{fan_temporal_2003}
Shanhui Fan, Wonjoo Suh, and J.~D. Joannopoulos.
\newblock Temporal coupled-mode theory for the {Fano} resonance in optical
  resonators.
\newblock {\em J. Opt. Soc. Am. A, JOSAA}, 20(3):569--572, March 2003.

\bibitem{wonjoo_suh_temporal_2004}
{Wonjoo Suh}, {Zheng Wang}, and {Shanhui Fan}.
\newblock Temporal coupled-mode theory and the presence of non-orthogonal modes
  in lossless multimode cavities.
\newblock {\em IEEE Journal of Quantum Electronics}, 40(10):1511--1518, October
  2004.

\bibitem{kristensen_theory_2017}
Philip~Trost Kristensen, Jakob~Rosenkrantz de~Lasson, Mikkel Heuck, Niels
  Gregersen, and Jesper Mørk.
\newblock On the {Theory} of {Coupled} {Modes} in {Optical}
  {Cavity}-{Waveguide} {Structures}.
\newblock {\em Journal of Lightwave Technology}, 35(19):4247--4259, October
  2017.

\bibitem{tromborg_transmission_1987}
B.~Tromborg, H.~Olesen, Xing Pan, and S.~Saito.
\newblock Transmission line description of optical feedback and injection
  locking for {Fabry}-{Perot} and {DFB} lasers.
\newblock {\em IEEE Journal of Quantum Electronics}, 23(11):1875--1889,
  November 1987.

\bibitem{morse_methods_1953}
Philip~McCord Morse, Herman Feshbach, and G.~P. Harnwell.
\newblock {\em Methods of {Theoretical} {Physics}, {Part} {I}}.
\newblock McGraw-Hill Book Company, Boston, Mass, June 1953.

\bibitem{ibanez_adiabaticity_2014}
S.~Ib\'a\~nez and J.~G. Muga.
\newblock Adiabaticity condition for non-hermitian hamiltonians.
\newblock {\em Phys. Rev. A}, 89:033403, Mar 2014.

\bibitem{arfken_mathematical_2005}
George~B. Arfken and Hans~J. Weber.
\newblock {\em Mathematical {Methods} for {Physicists}, 6th {Edition}}.
\newblock Academic Press, Boston, 6th edition edition, July 2005.

\bibitem{dembowski_experimental_2001}
C.~Dembowski, H.-D. Gräf, H.~L. Harney, A.~Heine, W.~D. Heiss, H.~Rehfeld, and
  A.~Richter.
\newblock Experimental {Observation} of the {Topological} {Structure} of
  {Exceptional} {Points}.
\newblock {\em Phys. Rev. Lett.}, 86(5):787--790, January 2001.

\bibitem{heiss_exceptional_2004}
W.D. Heiss.
\newblock Exceptional {Points} – {Their} {Universal} {Occurrence} and {Their}
  {Physical} {Significance}.
\newblock {\em Czechoslovak Journal of Physics}, 54(10):1091--1099, October
  2004.

\bibitem{berry_physics_2004}
M.V. Berry.
\newblock Physics of {Nonhermitian} {Degeneracies}.
\newblock {\em Czechoslovak Journal of Physics}, 54(10):1039--1047, October
  2004.

\bibitem{heiss_physics_2012}
W~D Heiss.
\newblock The physics of exceptional points.
\newblock {\em Journal of Physics A: Mathematical and Theoretical},
  45(44):444016, November 2012.

\bibitem{liertzer_pump-induced_2012}
M.~Liertzer, Li~Ge, A.~Cerjan, A.~D. Stone, H.~E. T\"ureci, and S.~Rotter.
\newblock Pump-induced exceptional points in lasers.
\newblock {\em Phys. Rev. Lett.}, 108:173901, Apr 2012.

\bibitem{bender_generalized_2002}
Carl~M Bender, M~V Berry, and Aikaterini Mandilara.
\newblock Generalized {PT} symmetry and real spectra.
\newblock {\em Journal of Physics A: Mathematical and General},
  35(31):L467--L471, August 2002.

\bibitem{ruter_observation_2010}
Christian~E. Rüter, Konstantinos~G. Makris, Ramy El-Ganainy, Demetrios~N.
  Christodoulides, Mordechai Segev, and Detlef Kip.
\newblock Observation of parity–time symmetry in optics.
\newblock {\em Nat. Phys.}, 6(3):192--195, March 2010.

\bibitem{feng_non-hermitian_2017}
Liang Feng, Ramy El-Ganainy, and Li~Ge.
\newblock Non-{Hermitian} photonics based on parity–time symmetry.
\newblock {\em Nat. Photonics}, 11(12):752--762, December 2017.

\bibitem{el-ganainy_non-hermitian_2018}
Ramy El-Ganainy, Konstantinos~G. Makris, Mercedeh Khajavikhan, Ziad~H.
  Musslimani, Stefan Rotter, and Demetrios~N. Christodoulides.
\newblock Non-{Hermitian} physics and {PT} symmetry.
\newblock {\em Nat. Phys.}, 14(1):11--19, January 2018.

\bibitem{klaiman_visualization_2008}
Shachar Klaiman, Uwe G\"unther, and Nimrod Moiseyev.
\newblock Visualization of branch points in $\mathcal{P}\mathcal{T}$-symmetric
  waveguides.
\newblock {\em Phys. Rev. Lett.}, 101:080402, Aug 2008.

\bibitem{berry_mode_nodate}
M~V Berry.
\newblock Mode degeneracies and the {Petermann} excess-noise factor for
  unstable lasers.
\newblock {\em J. Mod. Opt.}, 50(1):63--81, 2003.

\bibitem{kim_direct_2016}
Kyoung-Ho Kim, Min-Soo Hwang, Ha-Reem Kim, Jae-Hyuck Choi, You-Shin No, and
  Hong-Gyu Park.
\newblock Direct observation of exceptional points in coupled photonic-crystal
  lasers with asymmetric optical gains.
\newblock {\em Nat. Commun.}, 7:13893, December 2016.

\bibitem{lefebvre_resonance_2009}
R.~Lefebvre, O.~Atabek, M.~\ifmmode~\check{S}\else \v{S}\fi{}indelka, and
  N.~Moiseyev.
\newblock Resonance coalescence in molecular photodissociation.
\newblock {\em Phys. Rev. Lett.}, 103:123003, Sep 2009.

\bibitem{stehmann_observation_2004}
T~Stehmann, W~D Heiss, and F~G Scholtz.
\newblock Observation of exceptional points in electronic circuits.
\newblock {\em Journal of Physics A: Mathematical and General},
  37(31):7813--7819, August 2004.

\bibitem{xu_topological_2016}
H.~Xu, D.~Mason, Luyao Jiang, and J.~G.~E. Harris.
\newblock Topological energy transfer in an optomechanical system with
  exceptional points.
\newblock {\em Nature}, 537(7618):80--83, September 2016.

\bibitem{heiss_repulsion_2000}
W.~D. Heiss.
\newblock Repulsion of resonance states and exceptional points.
\newblock {\em Phys. Rev. E}, 61(1):929--932, January 2000.

\bibitem{rotter_non-hermitian_2009}
Ingrid Rotter.
\newblock A non-{Hermitian} {Hamilton} operator and the physics of open quantum
  systems.
\newblock {\em Journal of Physics A: Mathematical and Theoretical},
  42(15):153001, April 2009.

\bibitem{chong_symmetry_2012}
Chong, Y. D. and Ge, Li and Stone, A. Douglas.
\newblock $\mathcal{P}\mathcal{T}$-Symmetry Breaking and Laser-Absorber Modes in Optical Scattering Systems.
\newblock \emph{Phys. Rev. Lett.}, 108(26):269902(E), June 2012.

\bibitem{hanson_exceptional_2004}
G. W. Hanson, A. B. Yakovlev, M. A. K. Othman and F. Capolino.
\newblock Exceptional Points of Degeneracy and Branch Points for Coupled Transmission Lines—Linear-Algebra and Bifurcation Theory Perspectives.
\newblock \emph{IEEE Transactions on Antennas and Propagation}, 67(2):1025--1034, February 2019.

\bibitem{bandelow_theory_1993}
U.~Bandelow, H.J. Wunsche, and H.~Wenzel.
\newblock Theory of selfpulsations in two-section {DFB} lasers.
\newblock {\em IEEE Photonics Technology Letters}, 5(10):1176--1179, October
  1993.

\bibitem{wenzel_mechanisms_1996}
H.~Wenzel, U.~Bandelow, H.-J. Wunsche, and J.~Rehberg.
\newblock Mechanisms of fast self pulsations in two-section {DFB} lasers.
\newblock {\em IEEE Journal of Quantum Electronics}, 32(1):69--78, January
  1996.

\bibitem{feng_single-mode_2014}
L.~Feng, Z.~J. Wong, R.-M. Ma, Y.~Wang, and X.~Zhang.
\newblock Single-mode laser by parity-time symmetry breaking.
\newblock {\em Science}, 346(6212):972--975, November 2014.

\bibitem{hodaei_parity-time-symmetric_2014}
H.~Hodaei, M.-A. Miri, M.~Heinrich, D.~N. Christodoulides, and M.~Khajavikhan.
\newblock Parity-time-symmetric microring lasers.
\newblock {\em Science}, 346(6212):975--978, November 2014.

\bibitem{guckenheimer_nonlinear_1983}
John Guckenheimer and Philip Holmes.
\newblock {\em Nonlinear oscillations, dynamical systems, and bifurcations of vector fields}.
\newblock Springer-Verlag, New York, October 1983.


\bibitem{gunther_projective_2007}
Uwe Günther, Ingrid Rotter, and Boris~F Samsonov.
\newblock Projective {Hilbert} space structures at exceptional points.
\newblock {\em Journal of Physics A: Mathematical and Theoretical},
  40(30):8815--8833, July 2007.

\bibitem{keck_unfolding_2003}
F~Keck, H~J Korsch, and S~Mossmann.
\newblock Unfolding a diabolic point: a generalized crossing scenario.
\newblock {\em Journal of Physics A: Mathematical and General},
  36(8):2125--2137, February 2003.

\bibitem{muller_exceptional_2008}
Markus Müller and Ingrid Rotter.
\newblock Exceptional points in open quantum systems.
\newblock {\em Journal of Physics A: Mathematical and Theoretical},
  41(24):244018, June 2008.

\bibitem{menke_state_2016}
Henri Menke, Marcel Klett, Holger Cartarius, J\"org Main, and G\"unter Wunner.
\newblock State flip at exceptional points in atomic spectra.
\newblock {\em Phys. Rev. A}, 93:013401, Jan 2016.

\bibitem{heiss_collectivity_1998}
W.~D. Heiss, M.~Müller, and I.~Rotter.
\newblock Collectivity, phase transitions, and exceptional points in open
  quantum systems.
\newblock {\em Physical Review E}, 58(3):2894--2901, September 1998.

\bibitem{heiss_phases_1999}
W.D. Heiss.
\newblock Phases of wave functions and level repulsion.
\newblock {\em The European Physical Journal D - Atomic, Molecular and Optical
  Physics}, 7(1):1--4, August 1999.

\bibitem{glendinning_stability_1994}
Paul Glendinning.
\newblock {\em Stability, {Instability} and {Chaos}: {An} {Introduction} to the
  {Theory} of {Nonlinear} {Differential} {Equations}}.
\newblock Cambridge University Press, Cambridge England ; New York, 1 edition
  edition, November 1994.

\bibitem{iooss_elementary_1997}
Gerard Iooss and Daniel~D. Joseph.
\newblock {\em Elementary {Stability} and {Bifurcation} {Theory}}.
\newblock Springer, New York, 2nd edition edition, October 1997.

\bibitem{seydel_practical_2010}
R.~Seydel.
\newblock {\em Practical bifurcation and stability analysis}.
\newblock Number~5 in Interdisciplinary applied mathematics. Springer, New
  York, 3rd ed edition, 2010.
\newblock OCLC: ocn462919396.

\end{thebibliography}
\end{document}